\documentclass[12pt]{article}

\usepackage{amsmath}
\usepackage{cite}

\def\ti{\tilde}
\def\h{\hat}

\numberwithin{equation}{section}

\title{Lagrangian description of massive higher spin\\
supermultiplets in $AdS_3$ space}

\author{I.L. Buchbinder${}^{ab}$\thanks{joseph@tspu.edu.ru},
T.V. Snegirev${}^{ac}$\thanks{snegirev@tspu.edu.ru}, Yu.M.
Zinoviev${}^{de}$\thanks{Yurii.Zinoviev@ihep.ru}
\\[0.5cm]
\it ${}^a$Department of Theoretical Physics,\\
\it Tomsk State Pedagogical University,\\
\it Tomsk 634061, Russia\\[0.3cm]
\it ${}^b$National Research Tomsk State University,\\
 Tomsk 634050, Russia\\[0.3cm]
\it ${}^c$National Research Tomsk Polytechnic University,\\
Tomsk 634050, Russia\\[0.3cm]
\it ${}^d$Institute for High Energy Physics,\\
\it of National Research Center "Kurchatov Institute" \\
\it Protvino, Moscow Region, 142281, Russia \\[0.3cm]
\it ${}^e$Moscow Institute of Physics and Technology (State
University), \\
\it Dolgoprudny, Moscow Region, 141701, Russia}

\date{}

\begin{document}

\maketitle

\begin{abstract}
We construct the Lagrangian formulation of massive higher spin
on-shell (1,0) supermultiplets in three dimensional anti-de Sitter
space. The construction is based on description of the massive three
dimensional fields in terms of frame-like gauge invariant formalism
and technique of gauge invariant curvatures. For the two possible
massive supermultiplets $(s,s+1/2)$ and $(s,s-1/2)$ we derive
explicit form  of the supertransformations leaving the sum of
bosonic and fermionic Lagrangians invariant.
\end{abstract}

\thispagestyle{empty}
\newpage
\setcounter{page}{1}

\section{Introduction}

Three dimensional field models attract much attention due to their
comparatively simple structure and remarkable properties of
three-dimensional flat and curved spaces. One of the achievements in
this area is a construction of three-dimensional higher spin field
theories (see the pioneer works \cite{Blenc}, \cite{V1}, \cite{V2},
for modern development see e.g. the recent papers \cite{Co},
\cite{Bo} and references therein).

The aim of this work is to construct the massive higher spin
supermultiplets in the three-dimensional anti-de Sitter space
$AdS_3$. As is well known \cite{AT86}, the $AdS_3$ space possesses
the special properties since all the $AdS_3$ superalgebras (as well
as conventional $AdS_3$ algebra itself) factorize into "left" and
"right" parts. For the case of simplest (1,0) superalgebra (the one
we are working here with) such factorization has the form:
$$
OSp(1,2) \otimes Sp(2)
$$
so that we have supersymmetry in the "left" sector only. It means
that the minimal massive supermultiplet must contain just one
bosonic and one fermionic degrees of freedom. Recently we
constructed such supermultiplets using the unfolded formalism
\cite{BSZ16}. In this paper we develop the Lagrangian formulation
for these supermultiplets. It generalizes the Lagrangian formulation
for massive supermultiplets in three dimensional Minkowski space
given in \cite{BSZ15}. Note here that the off-shell superfield
description for the supermultiplets containing topologically massive
higher spin fields was constructed recently in
\cite{KT16}\footnote{The off-shell $3D, {\cal N}=2$ massless higher
spin superfields and their massive deformations has been elaborated
in \cite{KO}. The conditions defying the ${\cal N}=1$ massive
superfield representations in $AdS_3$ were formulated in
\cite{KU}.}.

We develop the component approach\footnote{List of references on
component and superfield formulations of higher spin supersymmetric
theories is given in our paper \cite{BSZ16}.} to Lagrangian
construction for supersymmetric massive higher spin fields in
$AdS_3$ on the base of gauge invariant formulation of massive higher
spin fields. It was shown long enough \cite{Zin01,Met06}
that massive bosonic (fermionic) spin-$s$ field can be treated as
system of massless fields with spins $s,s-1,s-2,...0(1/2)$ coupled
by the Stueckelberg symmetries. Later, a frame-like formulation of
massive higher spin fields has been developed in the framework of
the same approach \cite{Zin08}. It allows us to reformulate the
massive higher spin theory in terms of gauge invariant objects
(curvatures). In massless four-dimensional higher spin theory such
curvatures are very convenient objects and allow, e.g, to built the
cubic higher spin interactions \cite{FV87,FV87a,Vas11,BPS12}.
However the construction of curvatures proposed in
\cite{FV87,FV87a,Vas11,BPS12} is essentially adapted only for
theories in four and higher dimensions. Nevertheless the formalism
of gauge invariant objects can be successfully applied to massive
higher spin fields in three dimensions as well. It was shown that in
three dimensions the gauge invariant Lagrangians for massive higher
spin fields \cite{BSZ12a,BSZ14a} can be rewritten in explicitly
gauge invariant form \cite{Zin16}. In this work we elaborate this
formalism for Lagrangian construction of massive higher spin
supermultiplets with minimal (1,0) supersymmetry where the algebra
of the supercharges has the form
$$
\{Q_\alpha,Q_\beta\}\sim
P_{\alpha\beta}+\frac{\lambda}{2}M_{\alpha\beta}
$$
Here $P_{\alpha(2)}$ and $M_{\alpha(2)}$ are the generators of
$AdS_3$ algebra with the following commutation relations (the
conventions and notations for the indices are specified in the end
of introduction)
$$
[M_{\alpha(2)},M_{\beta(2)}]\sim\varepsilon_{\alpha\beta}
M_{\alpha\beta},\qquad
[P_{\alpha(2)},P_{\beta(2)}]\sim\lambda^2\varepsilon_{\alpha\beta}M_{\alpha\beta}
$$
$$
[M_{\alpha(2)},P_{\beta(2)}]\sim\varepsilon_{\alpha\beta}
P_{\alpha\beta}
$$
Note that as in all three-dimensional $AdS$ supergravities \cite{AT86}
(see also \cite{Zan05} for higher dimensions) in our frame-like
formalism we consider supertransformations just as the part of local
superalgebra acting in the fiber only. As a result the commutators of
our supertransformations close off-shell without any need of some
auxiliary fields (see Section 6). In the massive supermultiplets case
the price we have to pay is that the Lagrangians are invariant under
the supertransformations up to the terms proportional to the spin-1
and spin-0 auxiliary fields equations only.

A general scheme for Lagrangian formulation of massive higher spin
supermultiplets looks as follows. Let $\Omega^A,\Phi^A$ is a set of
fields and $R^A,F^A$ is a set of curvatures corresponding to
frame-like gauge invariant formulation of massive bosonic and
fermionic fields respectively. The curvatures have the following
structure
$$
R^A=D\Omega^A+(e\Omega)^A,\qquad F^A=D\Phi^A+(e\Phi)^A
$$
where $e\equiv e_\mu{}^{\alpha\beta}$ is a non-dynamical background
$AdS_3$ frame. Lagrangian both for bosonic and for fermionic massive
higher spins are presented as the quadratic forms in curvatures
expression \cite{Zin16}
\begin{eqnarray}\label{Scheme}
{\cal{L}_{B}}=\sum R^A R^A,\qquad {\cal{L}_{F}}=\sum F^AF^A
\end{eqnarray}
where ${\cal L}_{B}$ is a bosonic field Lagrangian and ${\cal
L}_{F}$ is a fermionic one. Note that for massless higher spin
fields in three dimensions such form of the Lagrangians is
impossible. Then, in order to realize supersymmetry between
the bosonic and fermionic massive fields we deform the curvatures by
gravitino field $\Psi_\mu{}^\alpha$ with parameter of
supertransformations $\zeta^\alpha$. In the case of global
supersymmetry we consider gravitino field as a non-dynamical
background and parameter of supertransformations as global (i.e.
covariantly constant). Such a construction can be interpreted as a
supersymmetric theory in terms of background fields of supergravity.
Schematically the curvature deformations are written as
$$
\Delta R^A=(\Psi\Phi)^A,\qquad \Delta F^A=(\Psi\Omega)^A
$$
and supertransformations have the form
$$
\delta\Omega^A\sim(\Phi\zeta)^A,\qquad
\delta\Phi^A\sim(\Omega\zeta)^A
$$
The restrictions on the form of deformations and
supertransformations are imposed by requirement of covariant
transformations for deformed curvatures $\hat R^A=R^A+\Delta R^A$
$$
\delta\hat{R}^A\sim (F\zeta)^A,\qquad \delta\hat F^A\sim (R\zeta)^A
$$
Finally  the supersymmetric Lagrangian for given supermultiplets is
the sum of Lagrangians (\ref{Scheme}) where initial curvatures are
replaced by deformed ones $R,F\rightarrow\h R,\h F$. Possible
arbitrariness is fixed by the condition that the Lagrangian must be
invariant under the supertransformations:
$$
\delta\hat{\cal{L}}=\sum [\hat R^A\delta \hat R^A+\hat F^A\delta
\hat F^A]=\sum R^A(F\zeta)^A=0
$$
Aim of the current paper is to demonstrate how such a general scheme
can actually be realized for supermultiplets in $AdS_3$.

The paper is organized as follows. First of all we fix the notations
and conventions. In section 2, following the above scheme we
consider as an example, the detailed Lagrangian construction of
massive supermultiplet $(2,\frac32)$. Such an example illustrates all
the key points of the construction. The next sections are devoted to
generalizations of these results for arbitrary massive
supermultiplets. In section 3 we consider the gauge invariant
formulation of massive fields with spin $s$ and spin $s+\frac12$
which will be used for supersymmetric constructions. In particular,
we write out all field variables and gauge-invariant curvatures and
consider the Lagrangian in terms of these curvatures. Further,
following the above general scheme, we study two massive higher spin
supermultiplets. Supermultiplets $(s,s+\frac12)$ is studied in
section 4 and $(s,s+\frac12)$ is studied in section 5. At last, in
Section 6 we show how the $AdS_3$ superalgebra is realized on our
supermultiplets. In conclusion we summaries and discuss the results
obtained. For completeness, we include in the paper two appendices
devoted to frame-like description of massless fields in three
dimensions. Appendix A contains the bosonic fields, Appendix B
contains the fermionic ones.

\noindent {\bf Notations and conventions.} We use a frame-like
multispinor formalism where all the objects (3,2,1,0-forms) have
totally symmetric local spinor indices. To simplify the expressions
we will use the condensed notations for the spinor indices such that
e.g.
$$
\Omega^{\alpha(2k)} = \Omega^{(\alpha_1\alpha_2 \dots \alpha_{2k})}
$$
Also we will always assume that spinor indices denoted by the same
letters and placed on the same level are symmetrized, e.g.
$$
\Omega^{\alpha(2k)} \zeta^\alpha = \Omega^{(\alpha_1\dots \alpha_{2k}}
\zeta^{\alpha_{2k+1})}
$$
$AdS_3$ space will be described by the background frame (one-form)
$e^{\alpha(2)}$ and the covariant derivative $D$ normalized so that
$$
D \wedge D \zeta^\alpha = - \lambda^2 E^\alpha{}_\beta \zeta^\beta
$$
Basis elements of $1,2,3$-form spaces are respectively
$e^{\alpha(2)}$, $E{}^{\alpha(2)}$, $E$ where the last two are
defined as double and triple wedge product of $e^{\alpha(2)}$:
$$
e^{\alpha\alpha} \wedge e^{\beta\beta} =
\varepsilon^{\alpha\beta}{E}{}^{\alpha\beta}, \qquad
E{}^{\alpha\alpha} \wedge e^{\beta\beta} = \varepsilon^{\alpha\beta}
\varepsilon^{\alpha\beta} E.
$$
Also we write some useful relations for these basis elements
$$
E{}^\alpha{}_\gamma \wedge e^{\gamma\beta} = 3
\varepsilon^{\alpha\beta} E, \qquad e^\alpha{}_\gamma \wedge
e^{\gamma\beta} = 4 E{}^{\alpha\beta}.
$$
Further on the sign of wedge product $\wedge$ will be omitted.

\section{Massive supermultiplet $(2,\frac32)$ example}

In this section we consider in details the Lagrangian realization
for massive supermultiplet $(2,3/2)$ example using the method
mentioned above in the introduction. We start with the gauge-invariant
formulations of free massive fields with spin 2 and spin 3/2
separately. Using such formulations, we present the full set of
gauge-invariant curvatures for them. Then we consider the
deformations of these curvatures by background gravitino field
$\Psi^\alpha$ and find suitable supertransformations. As a result we
construct the supersymmetric Lagrangian.

\subsection{Free fields}

{\bf Spin 2}
\\
In gauge invariant form the massive spin-2 field is described by
system of massless fields with spins $2,1,0$. In frame-like approach
the corresponding set of fields consists of
$(\Omega^{\alpha(2)},f^{\alpha(2)})$, $(B^{\alpha(2)},A)$ and
$(\pi^{\alpha(2)},\varphi)$ (see Appendix A for details). Lagrangian
for free massive field in $AdS_3$ have the form \cite{BSZ12a}
\begin{eqnarray}\label{Lag2}
{\cal{L}}&=&\Omega_{\alpha\beta}e^\beta{}_\gamma\Omega^{\alpha\gamma}+
\Omega_{\alpha(2)}Df^{\alpha(2)}+
EB^{\alpha(2)}B_{\alpha(2)}\nonumber\\
&&-e^{\alpha(2)}B_{\alpha(2)}DA-
E\pi^{\alpha(2)}\pi_{\alpha(2)}+E^{\alpha(2)}\pi_{\alpha(2)}D\varphi\nonumber\\
&&+ 2me_{\alpha(2)}\Omega^{\alpha(2)}A+m
f_{\alpha\beta}E^{\beta}{}_\gamma
B^{\alpha\gamma}+4\tilde{m}E_{\alpha(2)}\pi^{\alpha(2)}A
\nonumber\\
&&+\frac{M^2}{4}f_{\alpha\beta}e^\beta{}_\gamma f^{\alpha\gamma}
-m\tilde{m}E_{\alpha(2)}f^{\alpha(2)}\varphi+\frac32m{}^2E\varphi\varphi
\end{eqnarray}
It is invariant under the following gauge transformations
\begin{align}\label{GT2}
&
\delta\Omega^{\alpha(2)}=D\eta^{\alpha(2)}+\frac{M^2}{4}e^\alpha{}_\beta\xi^{\alpha\beta}
&& \delta
f^{\alpha(2)}=D\xi^{\alpha(2)}+e^\alpha{}_\beta\eta^{\alpha\beta}-2me^{\alpha(2)}\xi
\nonumber\\
& \delta B^{\alpha(2)}=-2m\eta^{\alpha(2)} && \delta
A=D\xi-\frac{m}{4}e_{\alpha(2)}\xi^{\alpha(2)}
\nonumber\\
&\delta\pi^{\alpha(2)}=-m\tilde{m}\xi^{\alpha(2)} &&
\delta\varphi=-4\tilde{m}\xi
\end{align}
where $m$ is the mass parameter and
$$
\tilde{m}^2=M^2,\quad M^2=m^2+\lambda^2
$$
The curvatures invariant under gauge transformations (\ref{GT2})
have the form
\begin{eqnarray}\label{Curv2}
{R}^{\alpha(2)}&=&D\Omega^{\alpha(2)}+\frac{M^2}{4}e^\alpha{}_\beta
f^{\alpha\beta}-{m}^{2}E^\alpha{}_\beta
B^{\alpha\beta}-m\tilde{m}E^{\alpha(2)}\varphi
\nonumber\\
T^{\alpha(2)}&=&Df^{\alpha(2)}+e^\alpha{}_\beta\Omega^{\alpha\beta}-2me^{\alpha(2)}A
\nonumber\\
{\cal{B}}^{\alpha(2)}&=&DB^{\alpha(2)}-\Omega^{\alpha(2)}+\frac{M^2}{4}e^\alpha{}_\beta\pi^{\alpha\beta}
\nonumber\\
{\cal{A}}&=&DA-\frac{m}{4}e_{\alpha(2)}f^{\alpha(2)}+2mE_{\alpha(2)}
B^{\alpha(2)}
\nonumber\\
\Pi^{\alpha(2)}&=&D\pi^{\alpha(2)}-f^{\alpha(2)}+e^\alpha{}_\beta
B^{\alpha\beta}+\frac{m}{2M}e^{\alpha(2)}\varphi
\nonumber\\
\Phi&=&D\varphi+4\tilde{m}A+mMe_{\alpha(2)}\pi^{\alpha(2)}
\end{eqnarray}
Here we have changed a normalization for the two zero forms
\begin{eqnarray}\label{Norm2}
B^{\alpha(2)}\rightarrow -2mB^{\alpha(2)}\qquad
\pi^{\alpha(2)}\rightarrow -mM\pi^{\alpha(2)}
\end{eqnarray}
It is interesting to point out that, unlike massless theory in three
dimensions, the  Lagrangian for massive spin 2 (\ref{Lag2}) can be
presented in manifestly gauge invariant form
\begin{eqnarray}\label{LagC2}
{\cal{L}}&=&-\frac{1}{2}{\cal{R}}_{\alpha(2)}\Pi^{\alpha(2)}
-\frac{1}{2}{\cal{T}}_{\alpha(2)}{\cal{B}}^{\alpha(2)}
-\frac{m}{4\tilde{m}}e_{\alpha(2)}{\cal{B}}^{\alpha(2)}\Phi
\end{eqnarray}
\\
\vspace{0.3cm}
\\
{\bf Spin-$\frac32$}
\\
In gauge invariant formulation, the  massive spin-3/2 field is
described by system of massless fields with the spins $3/2,1/2$. In
frame-like approach the corresponding set of fields consists of
$\Phi^\alpha$, $\phi^\alpha$. (see Appendix B for details). The free
Lagrangian for mass $m_1$ field in $AdS_3$ has the following form
\begin{eqnarray}\label{Lag3/2}
{\cal{L}}&=&-\frac{i}{2}\Phi_\alpha
D\Phi^\alpha+\frac{i}{2}\phi_\alpha E^\alpha{}_\beta
D\phi^\beta-\nonumber\\
&&-i\dfrac{M_1}{2}\Phi_\alpha
e^\alpha{}_\beta\Phi^\beta-i2m_1\Phi_\alpha
E^\alpha{}_\beta\phi^\beta-i\dfrac{3M_1}{2}E\phi_\alpha\phi^\alpha
\end{eqnarray}
It is invariant under gauge transformations
\begin{eqnarray}\label{GT3/2}
\delta\Phi^\alpha=D\xi^\alpha+M_1e^\alpha{}_\beta\xi^\beta \qquad
\delta\phi^\alpha=-2m_1\xi^\alpha
\end{eqnarray}
where
$$
M_1{}^2=m_1{}^2+\frac{\lambda^2}{4}
$$
One can construct the gauge invariant curvatures with respect
(\ref{GT3/2}). After change of normalization
$\phi^{\alpha}\rightarrow -2m_1\phi^{\alpha}$ they look like
\begin{eqnarray}\label{Curv3/2}
{\cal{F}}^\alpha&=&D\Phi^\alpha+M_1e^\alpha{}_\beta\Phi^\beta-4m_1{}^2E^\alpha{}_\beta\phi^\beta
\nonumber\\
{\cal{C}}^\alpha&=&D\phi^\alpha-\Phi^\alpha+M_1e^\alpha{}_\beta\phi^\beta
\end{eqnarray}
As in the previous spin-2 case, one notes that Lagrangian
(\ref{Lag3/2}) can be rewritten in terms of curvatures
(\ref{Curv3/2})
\begin{eqnarray}\label{LagC3/2}
{\cal{L}}&=&\frac{i}{2}{\cal{F}}_{\alpha}{\cal{C}}^{\alpha}
\end{eqnarray}

The above results on spin-2 and spin-$\frac{3}{2}$ fields are
building blocks for supersymmetrization. Full set of gauge invariant
curvatures contains  (\ref{Curv2}) and (\ref{Curv3/2}) respectively.
Corresponding expressions for Lagrangians in terms of curvatures are
(\ref{LagC2}) and (\ref{LagC3/2}).

\subsection{Supersymmetric system}

Before we turn to Lagrangian formulation for massive supermultiplet
$(2,3/2)$ let us consider this supermultiplet in massless flat
limit. That is we consider the sum of Lagrangians (\ref{Curv2}) and
(\ref{Lag3/2}) in the limit $m,m_1,\lambda\rightarrow0$. It has the
form
\begin{eqnarray}\label{SLag2}
\h{\cal L} &=&  \Omega_{\alpha\beta} e^\beta{}_\gamma
\Omega^{\alpha\gamma} + \Omega_{\alpha(2)} D f^{\alpha(2)} + E
B_{\alpha\beta} B^{\alpha\beta} - B_{\alpha\beta} e^{\alpha\beta} D
A\nonumber\\
&& - E \pi_{\alpha\beta} \pi^{\alpha\beta} + \pi_{\alpha\beta}
E^{\alpha\beta} D \varphi -\frac{i}{2} \Phi_{\alpha} D \Phi^{\alpha}
+ \frac{1}{2} \phi_\alpha E^\alpha{}_\beta D \phi^\beta
\end{eqnarray}
Such a Lagrangian describes the system of free massless fields with
spins $2,3/2,1,1/2,0$. One can show there exist global
supertransformations that leave the Lagrangian (\ref{SLag2})
invariant. These supertransformations with the redefinition
(\ref{Norm2}) have form
\begin{eqnarray}\label{ST2}
\delta f^{\alpha(2)} &=&i\beta_1\Phi^{\alpha}\zeta^\alpha
\nonumber\\
\delta A &=&i\alpha_0\Phi^\alpha\zeta_\alpha
-i2m_1\beta_0e_{\alpha\beta}\phi^\alpha\zeta^\beta
\nonumber\\
\delta\varphi &=&im_1\tilde\delta_0\phi^\gamma\zeta_\gamma
\nonumber\\
\delta\Phi^{\alpha} &=&
2\beta_1\Omega^{\alpha\beta}\zeta_\beta-2m\alpha_0e_{\beta(2)}B^{\beta(2)}\zeta^\alpha
\nonumber\\
\delta\phi^\alpha&=&\frac{4m\beta_0}{m_1}B^{\alpha\beta}\zeta_\beta+
\frac{mM\ti\delta_0}{2m_1}\pi^{\alpha\beta}\zeta_\beta
\end{eqnarray}
The parameters $\beta_1,\beta_0,\alpha_0,\ti\delta_0$ are arbitrary
but, as it will be seen later, they are fixed in massive case. The
choice of such notations for them will be clear from in next section
for arbitrary higher spin supermultiplets. To prove the invariance
we need to use the Lagrangian equations of motion for auxiliary
fields $\pi^{\alpha(2)},B^{\alpha(2)}$ corresponding to spins 0 and
1
\begin{eqnarray}\label{Low_eq}
E^\alpha{}_\beta DB^{\beta\gamma}=E^\gamma{}_\beta DB^{\beta\alpha}
,\qquad E^\alpha{}_\gamma d \pi^{\beta\gamma}
=\frac12\varepsilon^{\alpha\beta}E_{\gamma\delta}D\pi^{\gamma\delta}
\end{eqnarray}

Thus, we see that in the massless flat limit the Lagrangian
formulation of pure massive supermultiplets $(2,\frac32)$ we must get
the supertransformations (\ref{ST2}). This requirement will provide us
the unique possibility for construction of the correct supersymmetric
massive theory.
\\
\\
{\bf Deformation of curvatures}
\\
Now we are ready to realize the massive supermultiplets. We do it
deforming curvatures by background gravitino field $\Psi^\alpha$
with global transformation in $AdS_3$
\begin{eqnarray}\label{Trans3/2}
\delta\Psi^\alpha=D\zeta^\alpha+\frac{\lambda}{2}e^\alpha{}_\beta\zeta^\beta
\end{eqnarray}
The main idea is to deform the curvatures so that they transform
covariantly through themselves. It means that for all the deformed
curvatures $\hat R^A=R^A+\Delta R^A$ the following relations should
take place
\begin{eqnarray}\label{CReq}
 \delta\hat R^A=\delta_\zeta
R^A+\delta_0\Delta R^A=(R\zeta)^A
\end{eqnarray}
where $\delta_0$ is the transformation (\ref{Trans3/2}) and
$\delta_\zeta$ is the linear supertransformation.

Let us consider the following ansatz for deformation of spin-2
curvatures
\begin{align}\label{CD2}
& \Delta{R}^{\alpha(2)}=i\rho_1\Phi^\alpha\Psi^\alpha-i\h\rho_0
e^{\alpha(2)}\phi_\gamma\Psi^\gamma && \Delta
T^{\alpha(2)}=i\beta_1\Phi^\alpha\Psi^\alpha
\nonumber\\
&\Delta{\cal{B}}^{\alpha(2)}=-i\h\rho_1\phi^\alpha\Psi^\alpha &&
\Delta{\cal{A}}=i\alpha_0\Phi^\gamma\Psi_\gamma-i2m_1\beta_0e_{\alpha\beta}\phi^\alpha\Psi^\beta
\nonumber\\
&\Delta\Pi^{\alpha(2)}=-i\h\beta_1\phi^\alpha\Psi^\alpha
&&\Delta\Phi=im_1\ti\delta_0\phi_\gamma\Psi^\gamma
\end{align}
Also one introduces the corresponding supertransformations
\begin{align}\label{MasST2}
& \delta\Omega^{\alpha(2)}=i\rho_1\Phi^\alpha\zeta^\alpha-i\h\rho_0
e^{\alpha(2)}\phi_\gamma\zeta^\gamma && \delta
f^{\alpha(2)}=i\beta_1\Phi^\alpha\zeta^\alpha
\nonumber\\
& \delta B^{\alpha(2)}=i\h\rho_1\phi^\alpha\zeta^\alpha && \delta
A=i\alpha_0\Phi^\gamma\zeta_\gamma-i2m_1\beta_0e_{\alpha\beta}\phi^\alpha\zeta^\beta
\nonumber\\
&\delta\pi^{\alpha(2)}=i\h\beta_1\phi^\alpha\zeta^\alpha &&
\delta\varphi=-im_1\tilde\delta_0\phi_\gamma\zeta^\gamma
\end{align}
The form of the supertransformations is completely determined by the
form of deformations for curvatures. Besides
$\beta_1,\beta_0,\alpha_0,\ti\delta_0$ parameters which remain
arbitrary from the massless case, we have new arbitrary parameters
$\rho_1,\h\rho_1,\h\rho_0,\h\beta_1$. All the parameters will be
fixed by requirement of covariant homogeneous curvature
transformation (\ref{CReq}). In addition there are two mass
parameters $M,M_1$ an we will see that one is related to another.

Let us check the requirement (\ref{CReq}) for curvatures
${\cal{R}}^{\alpha(2)},{\cal{T}}^{\alpha(2)}$. One one hand
\begin{eqnarray*}
\delta\hat{\cal{R}}^{\alpha(2)} &=&
i\rho_1D\Phi^{\alpha}\zeta^\alpha
+i\hat\rho_0e^{\alpha(2)}D\phi^\beta\zeta_\beta+
i\frac{M^2}{2}\beta_1e^{\alpha(2)} \Phi^{\beta}\zeta_\beta\\
&&+
i(\frac{M^2}{2}\beta_1-\frac12\lambda\rho_1)e^\alpha{}_\gamma\Phi^{\alpha}\zeta^\gamma
+i(-m{}^2\h\rho_1-\frac12\lambda\h\rho_0
+\frac12mMm_1\ti\delta_0)E^\alpha{}_\gamma\phi^{\alpha}\zeta^\gamma\\
&&+ i(-m{}^2\h\rho_1-\frac12\lambda\h\rho_0
-\frac12mMm_1\ti\delta_0)E^\alpha{}_\gamma\phi^{\gamma}\zeta^\alpha
\\
\delta\hat{\cal{T}}^{\alpha(2)} &=&
i\beta_1D\Phi^{\alpha}\zeta^\alpha+
i(-2m\alpha_0+2\rho_1)e^{\alpha(2)}
\Phi^{\beta}\zeta_\beta+i(2\rho_1-\frac12\lambda\beta_1)e^\alpha{}_\gamma\Phi^{\alpha}\zeta^\gamma\\
&&+
i(-4mm_1\beta_0-4\h\rho_0)E^\alpha{}_\gamma\phi^{\alpha}\zeta^\gamma+
i(-4mm_1\beta_0+4\h\rho_0)E^\alpha{}_\gamma\phi^{\gamma}\zeta^\alpha
\end{eqnarray*}
and on the other hand
\begin{eqnarray}\label{CT_1}
\delta\hat{\cal{R}}^{\alpha(2)} &=&
i\rho_1{\cal{F}}^{\alpha}\zeta^\alpha
-i\hat\rho_0e^{\alpha(2)}{\cal{C}}^\beta\zeta_\beta=i\rho_1D\Phi^{\alpha}\zeta^\alpha
+i\hat\rho_0e^{\alpha(2)}D\phi^\beta\zeta_\beta\nonumber\\
&&+ i(\h\rho_0+2M_1\rho_1)e^{\alpha(2)} \Phi^{\beta}\zeta_\beta+
i(M_1\rho_1)e^\alpha{}_\gamma\Phi^{\alpha}\zeta^\gamma\nonumber\\
&&-iM_1\h\rho_0E^\alpha{}_\gamma\phi^{\alpha}\zeta^\gamma+ i(
-M_1\h\rho_0-4m_1{}^2\rho_1)E^\alpha{}_\gamma\phi^{\gamma}\zeta^\alpha
\nonumber\\
\delta\hat{\cal{T}}^{\alpha(2)} &=&
i\beta_1{\cal{F}}^{\alpha}\zeta^\alpha
=i\beta_1D\Phi^{\alpha}\zeta^\alpha+ i(2M_1\beta_1)e^{\alpha(2)}
\Phi^{\beta}\zeta_\beta\nonumber\\
&&+i(M_1\beta_1)e^\alpha{}_\gamma\Phi^{\alpha}\zeta^\gamma
-i4m_1{}^2\beta_1E^\alpha{}_\gamma\phi^{\gamma}\zeta^\alpha
\end{eqnarray}
Comparing the above relations, we obtain the solution
$$
M_1=M-\frac{\lambda}{2},\quad\rho_1=\frac{M}{2}\beta_1,\quad\h\rho_0=-\frac{M(M-\lambda)}{2}\beta_1,\quad
\alpha_0=-\frac{(M-\lambda)}{2m}\beta_1
$$
$$
\ti\delta_0=4\beta_0=\frac{2m_1}{m}\beta_1
$$
For curvatures ${\cal{B}}^{\alpha(2)},\Pi^{\alpha(2)}$ we have on
one hand
\begin{eqnarray*}
\delta\hat{\cal{B}}^{\alpha(2)} &=&
i\hat\rho_1D\phi^\alpha\zeta^\alpha
-i\rho_1\Phi^{\alpha}\zeta^\alpha\\
&&+ i(-\h\rho_0+\frac{M^2}{2}\h\beta_1)e^{\alpha(2)}
\phi^{\beta}\zeta_\beta+
i(\frac{M^2}{2}\h\beta_1-\frac12\lambda\h\rho_1)e^\alpha{}_\gamma\phi^{\alpha}\zeta^\gamma
\\
\delta\hat\Pi^{\alpha(2)} &=&i\hat\beta_1D\phi^\alpha\zeta^\alpha
-i\beta_1\Phi^{\alpha}\zeta^\alpha
\\
&&+ i(\frac{mm_1\ti\delta_0}{2M}+2\h\rho_1)e^{\alpha(2)}
\phi^{\beta}\zeta_\beta+i(2\h\rho_1-\frac12\lambda\h\beta_1)e^\alpha{}_\gamma\phi^{\alpha}\zeta^\gamma
\end{eqnarray*}
and on the other hand
\begin{eqnarray}\label{CT_2}
\delta\hat{\cal{B}}^{\alpha(2)} &=&
i\hat\rho_1{\cal{C}}^\alpha\zeta^\alpha=i\hat\rho_1D\phi^\alpha\zeta^\alpha
-i\h\rho_1\Phi^{\alpha}\zeta^\alpha\nonumber\\
&&+ i(2M_1\h\rho_1)e^{\alpha(2)} \phi^{\beta}\zeta_\beta+
i(M_1\h\rho_1)e^\alpha{}_\gamma\phi^{\alpha}\zeta^\gamma
\nonumber\\
\delta\hat\Pi^{\alpha(2)}
&=&i\hat\beta_1{\cal{C}}^\alpha\zeta^\alpha=i\hat\beta_1D\phi^\alpha\zeta^\alpha
-i\h\beta_1\Phi^{\alpha}\zeta^\alpha\nonumber\\
&&+ i(2M_1\h\beta_1)e^{\alpha(2)}
\phi^{\beta}\zeta_\beta+i(M_1\h\beta_1)e^\alpha{}_\gamma\phi^{\alpha}\zeta^\gamma
\end{eqnarray}
Comparison of above relations yield
$$
\h\beta_1=\beta_1,\qquad\h\rho_1=\rho_1
$$
The transformation laws for curvatures ${\cal{A}}$ and $\Phi$ look
like
\begin{eqnarray}\label{CT_3}
\delta\hat{\cal{A}} = i\alpha_0{\cal{F}}^\alpha\zeta_\alpha
+i2m_1\beta_0e_{\alpha\beta}{\cal{C}}^\alpha\zeta^\beta,\quad
\delta\hat\Phi =im_1\tilde\delta_0{\cal{C}}^\gamma\zeta_\gamma
\end{eqnarray}

Now we consider ansatz for deformation of spin-3/2 curvatures
\begin{eqnarray*}
\Delta{\cal{F}}^\alpha&=&2\beta_1\Omega^{\alpha\beta}\Psi_\beta-2m\alpha_0e_{\gamma(2)}B^{\gamma(2)}\Psi^\alpha
+\delta_0f^{\alpha\beta}\Psi_\beta+\gamma_0A\Psi^\alpha-\ti\gamma_0e^{\alpha\beta}\varphi\Psi_\beta
\\
\Delta{\cal{C}}^\alpha&=&-\frac{4m\beta_0}{m_1}B^{\alpha\beta}\Psi_\beta
-\frac{mM\ti\delta_0}{2m_1}\pi^{\alpha\beta}\Psi_\beta
-\rho_0\varphi\Psi^\alpha
\end{eqnarray*}
One introduces the corresponding supertransformations
\begin{eqnarray}\label{MasST3/2}
\delta\Phi^\alpha&=&2\beta_1\Omega^{\alpha\beta}\zeta_\beta-2m\alpha_0e_{\gamma(2)}B^{\gamma(2)}\zeta^\alpha
+\delta_0f^{\alpha\beta}\zeta_\beta+\gamma_0A\zeta^\alpha-\ti\gamma_0e^{\alpha\beta}\varphi\zeta_\beta
\nonumber\\
\delta\phi^\alpha&=&\frac{4m\beta_0}{m_1}B^{\alpha\beta}\zeta_\beta
+\frac{mM\ti\delta_0}{2m_1}\pi^{\alpha\beta}\zeta_\beta
+\rho_0\varphi\zeta^\alpha
\end{eqnarray}
Let us require the conditions (\ref{CReq}). The we have on one hand
\begin{eqnarray*}
\delta\hat{\cal{F}}^{\alpha}
&=&2\beta_{1}D\Omega^{\alpha\beta}\zeta_\beta+2m\alpha_0e_{\beta(2)}DB^{\beta(2)}\zeta^\alpha
+ \delta_0Df^{\alpha\beta}\zeta_\beta+\gamma_0DA\zeta^\alpha+
\tilde\gamma_0D\varphi e^\alpha{}_\beta\zeta^\beta\\
&&+(-\lambda\beta_1
-2M_1\beta_{1})e_{\gamma(2)}\Omega^{\alpha\gamma}\zeta^\gamma+
(2M_1\beta_{1})e_{\gamma(2)}\Omega^{\gamma(2)}\zeta^\alpha\\
&& +(-\frac12\lambda\delta_0
-M_1\delta_0)e_{\gamma(2)}f^{\alpha\gamma}\zeta^\gamma+
(M_1\delta_0)e_{\gamma(2)}f^{\gamma(2)}\zeta^\alpha
+(M_1\gamma_0-\frac12\lambda\gamma_0)e^\alpha{}_\gamma
A\zeta^\gamma\\
&&+(-4mM_1\alpha_0-16mm_1\beta_0+2m\lambda\alpha_0)E^\alpha{}_\gamma
B^{\gamma\beta}\zeta_\beta+(4mM_1\alpha_0-2m\lambda\alpha_0)E_{\gamma(2)}B^{\alpha\gamma}\zeta^\gamma\\
&&-2mMm_1\ti\delta_0E^\alpha{}_\gamma \pi^{\gamma\beta}\zeta_\beta
+(4M_1\ti\gamma_0-4m_1{}^2\rho_0+2\lambda\ti\gamma_0)E^\alpha{}_\gamma
\varphi\zeta^\gamma
\end{eqnarray*}
\begin{eqnarray*}
\delta\hat{\cal{C}}^\alpha&=&\frac{4m\beta_0}{m_1}DB^{\alpha\beta}\zeta_\beta
+\frac{mM\tilde\delta_0}{2m_1}D\pi^{\alpha\beta}\zeta_\beta
+\rho_0D\varphi\zeta^\alpha
-2\beta_{1}\Omega^{\alpha\beta}\zeta_\beta-
\delta_0f^{\alpha\beta}\zeta_\beta-\gamma_0A\zeta^\alpha\\
&&+(-\frac{2m\beta_0}{m_1}\lambda
-\frac{4mM_1\beta_0}{m_1})e_{\gamma(2)}B^{\alpha\gamma}\zeta^\gamma
+(2m\alpha_0+\frac{4mM_1\beta_0}{m_1})e_{\gamma(2)}B^{\gamma(2)}\zeta^\alpha\\
&&+ (-\frac{mM\tilde\delta_0}{4m_1}\lambda
-\frac{mMM_1\tilde\delta_0}{2m_1})e_{\gamma(2)}\pi^{\alpha\gamma}\zeta^\gamma\\
&&+(\frac{mMM_1\tilde\delta_0}{2m_1})e_{\gamma(2)}\pi^{\gamma(2)}\zeta^\alpha
+(-\tilde\gamma_0+M_1\rho_0- \frac12\lambda\rho_0)e^\alpha{}_\gamma
\varphi\zeta^\gamma
\end{eqnarray*}
and on the other hand
\begin{eqnarray}\label{CT_4}
\delta\hat{\cal{F}}^{\alpha}
&=&2\beta_{1}{\cal{R}}^{\alpha\beta}\zeta_\beta+2m\alpha_0e_{\beta(2)}{\cal{B}}^{\beta(2)}\zeta^\alpha
+
\delta_0{\cal{T}}^{\alpha\beta}\zeta_\beta+\gamma_0{\cal{A}}\zeta^\alpha+
\tilde\gamma_0\Phi e^\alpha{}_\beta\zeta^\beta\nonumber\\
&=&2\beta_{1}D\Omega^{\alpha\beta}\zeta_\beta+2m\alpha_0e_{\beta(2)}DB^{\beta(2)}\zeta^\alpha
+ \delta_0Df^{\alpha\beta}\zeta_\beta+\gamma_0DA\zeta^\alpha+
\tilde\gamma_0D\varphi e^\alpha{}_\beta\zeta^\beta\nonumber\\
&&-2\delta_0e_{\gamma(2)}\Omega^{\alpha\gamma}\zeta^\gamma+
(-2m\alpha_0+\delta_0)e_{\gamma(2)}\Omega^{\gamma(2)}\zeta^\alpha
-M^2\beta_{1}e_{\gamma(2)}f^{\alpha\gamma}\zeta^\gamma\nonumber\\
&&+
(-\frac14m\gamma_0+\frac{M^2}{2}\beta_{1})e_{\gamma(2)}f^{\gamma(2)}\zeta^\alpha
+(2m\delta_0-4M\tilde\gamma_0)e^\alpha{}_\gamma
A\zeta^\gamma\nonumber\\
&&+(-2m{}^2\beta_1+2m\gamma_0)E^\alpha{}_\gamma
B^{\gamma\beta}\zeta_\beta+(2m{}^2\beta_1+2m\gamma_0)E_{\gamma(2)}B^{\alpha\gamma}\zeta^\gamma\nonumber\\
&&+(-2mM\ti\gamma_0+4mM^2\alpha_0)E^\alpha{}_\gamma
\pi^{\gamma\beta}\zeta_\beta+(2mM\ti\gamma_0+4mM^2\alpha_0)E_{\gamma(2)}\pi^{\alpha\gamma}\zeta^\gamma\nonumber\\
&& +2mM\beta_1E^\alpha{}_\gamma \varphi\zeta^\gamma\
\end{eqnarray}
\begin{eqnarray}\label{CT_5}
\delta\hat{\cal{C}}^\alpha&=&\frac{4m\beta_0}{m_1}{\cal{B}}^{\alpha\beta}\zeta_\beta
+\frac{mM\ti\delta_0}{2m_1}\Pi^{\alpha\beta}\zeta_\beta
+\rho_0\Phi\zeta^\alpha\nonumber\\
&=& \frac{4m\beta_0}{m_1}DB^{\alpha\beta}\zeta_\beta
+\frac{mM\ti\delta_0}{2m_1}D\pi^{\alpha\beta}\zeta_\beta
+\rho_0D\varphi\zeta^\alpha
-\frac{4m\beta_0}{m_1}\Omega^{\alpha\beta}\zeta_\beta-
\frac{mM\ti\delta_0}{2m_1}f^{\alpha\beta}\zeta_\beta\nonumber\\
&&+4M\rho_0A\zeta^\alpha-\frac{mM\ti\delta_0}{m_1}e_{\gamma(2)}B^{\alpha\gamma}\zeta^\gamma
+\frac{mM\ti\delta_0}{2m_1}e_{\gamma(2)}B^{\gamma(2)}\zeta^\alpha
-\frac{2mM^2\beta_0}{m_1c_0}e_{\gamma(2)}\pi^{\alpha\gamma}\zeta^\gamma\nonumber\\
&&+(mM\rho_0+
\frac{mM^2\beta_0}{m_1})e_{\gamma(2)}\pi^{\gamma(2)}\zeta^\alpha
-\frac{m^2M\tilde\delta_0}{4m_1M}e^\alpha{}_\gamma
\varphi\zeta^\gamma
\end{eqnarray}
Comparison of the above relations gives us at
$M_1=M-\frac{\lambda}{2}$
$$
\delta_0=M\beta_1,\quad\gamma_0=-2\ti\gamma_0=-\frac{2M(M-\lambda)}{m}\beta_1,\quad
\rho_0=\frac{(M-\lambda)}{2m}\beta_1
$$

Thus imposing the condition of the covariant curvature
transformations (\ref{CReq}) we fix the supersymmetric deformations
up to common parameter $\beta_1$ and relation between mass
$M_1=M-\frac{\lambda}{2}$. The law of supersymmetry transformations
is given by (\ref{MasST2}),(\ref{MasST3/2}). The obtained result is
in agreement with our recent work \cite{BSZ16} where we studied the
analogous supermultiplets $(2,3/2)$ in unfolded formulation.
\\
\\
{\bf Invariant Lagrangian}
\\
Now we turn to construction of supersymmetric Lagrangian
corresponding to supermultiplets $(2,\frac23)$. Actually in terms of
curvatures it presents a sum of Lagrangians for spin 2 (\ref{LagC2})
and spin 3/2 (\ref{LagC3/2}), where the initial curvatures are
replaced by deformed ones $R\rightarrow\hat R$
\begin{eqnarray*}
\h{\cal{L}}&=&-\frac{1}{2}\h{\cal{R}}_{\alpha(2)}\h\Pi^{\alpha(2)}
-\frac{1}{2}\h{\cal{T}}_{\alpha(2)}\h{\cal{B}}^{\alpha(2)}
-\frac{m}{4\tilde{m}}e_{\alpha(2)}\h{\cal{B}}^{\alpha(2)}\h\Phi+
\frac{i}{2}\h{\cal{F}}_{\alpha}\h{\cal{C}}^{\alpha}
\end{eqnarray*}
Using the knowing supertransformations for curvatures
(\ref{CT_1}),(\ref{CT_2}),(\ref{CT_3}),(\ref{CT_4}),(\ref{CT_5}) it
is not very difficult to check the invariance of the Lagrangian.
Indeed
\begin{eqnarray*}
\delta\h{\cal{L}}&=&-i(\rho_1-\frac{mM}{4m_1}\ti\delta_0){\cal{F}}^{\alpha}\zeta^\alpha\Pi_{\alpha(2)}
-i(\hat\beta_1-\beta_1){\cal{R}}_{\alpha(2)}{\cal{C}}^\alpha\zeta^\alpha\\
&&
-i(\beta_1-\frac{2m}{m_1}\beta_0){\cal{F}}^{\alpha}\zeta^\alpha{\cal{B}}_{\alpha(2)}
-i(\hat\rho_1-\frac12\delta_0){\cal{T}}_{\alpha(2)}{\cal{C}}^\alpha\zeta^\alpha\\
&&
-i(\frac{mm_1}{4\tilde{m}}\tilde\delta_0+m\alpha_0)e_{\alpha(2)}\h{\cal{B}}^{\alpha(2)}{\cal{C}}^\gamma\zeta_\gamma
-\frac{i}{2}(\frac{m}{\tilde{m}}\hat\rho_1+\ti\gamma_0)e_{\alpha(2)}{\cal{C}}^\alpha\zeta^\alpha\Phi
\\
&&- \frac{i\gamma_0}{2}{\cal{A}}\zeta^\alpha{\cal{C}}_{\alpha}+
\frac{i\rho_0}{2}\h{\cal{F}}_{\alpha}\Phi\zeta^\alpha
+\frac{1}{2}i\hat\rho_0e^{\alpha(2)}{\cal{C}}^\beta\Pi_{\alpha(2)}\zeta_\beta
\end{eqnarray*}
Taking into account the equations of equation for fields
$B^{\alpha(2)},\pi^{\alpha(2)}$ which are equivalent to the
following relations
$$
\Phi=0,\quad{\cal{A}}=0\quad\Rightarrow\quad
e_{\gamma(2)}\Pi^{\gamma(2)}=D\Phi-4M{\cal{A}}=0
$$
we see that the variation $\delta\h{\cal{L}}$ vanishes.

\section{Free massive higher spin fields}
For completeness in this section we discuss the gauge invariant
formulation of free massive bosonic and fermionic higher spin fields
in $AdS_3$ \cite{BSZ12a,BSZ14a}. Besides, we present the gauge
invariant curvatures, the Lagrangians in explicit component form and
the Lagrangians in  terms of curvatures for bosonic spin $s$ and
fermionic spin $s+1/2$ fields.

\subsection{Bosonic spin-$s$ field}
In gauge invariant form the massive spin $s$ field is described by
system of massless fields with spins $s,(s-1),...,0$. In frame-like
approach the corresponding set of fields consists of
(\ref{massless_bosons})
$$
(\Omega^{\alpha(2k)},f^{\alpha(2k)})\quad1\leq k\leq (s-1),\quad
(B^{\alpha(2)},A),\quad (\pi^{\alpha(2)},\varphi)
$$
Free Lagrangian for the fields with mass $m$ in $AdS_3$ have the
form
\begin{eqnarray}\label{Lag_s}
{\cal L} &=& \sum_{k=1}^{s-1} (-1)^{k+1} [ k
\Omega_{\alpha(2k-1)\beta} e^\beta{}_\gamma
\Omega^{\alpha(2k-1)\gamma} + \Omega_{\alpha(2k)} D f^{\alpha(2k)}
 ] \nonumber \\
 && + E B_{\alpha\beta} B^{\alpha\beta} - B_{\alpha\beta}
e^{\alpha\beta} D A - E \pi_{\alpha\beta} \pi^{\alpha\beta} +
\pi_{\alpha\beta} E^{\alpha\beta} D \varphi \nonumber \\
 && + \sum_{k=1}^{s-2} (-1)^{k+1} a_k [ - \frac{(k+2)}{k}
\Omega_{\alpha(2)\beta(2k)} e^{\alpha(2)} f^{\beta(2k)} +
\Omega_{\alpha(2k)} e_{\beta(2)} f^{\alpha(2k)\beta(2)} ] \nonumber
\\
 && + 2a_0 \Omega_{\alpha(2)} e^{\alpha(2)} A - a_0
f_{\alpha\beta} E^\beta{}_\gamma B^{\alpha\gamma} + 2sM
\pi_{\alpha\beta} E^{\alpha\beta} A \nonumber \\
 && + \sum_{k=1}^{s-1} (-1)^{k+1} b_k f_{\alpha(2k-1)\beta}
e^\beta{}_\gamma f^{\alpha(2k-1)\gamma} + b_0 f_{\alpha(2)}
E^{\alpha(2)} \varphi + \frac{3a_0{}^2}{2}
 E \varphi^2
\end{eqnarray}
where
\begin{eqnarray}\label{boson_data}
&a_k{}^2 = \dfrac{k(s+k+1)(s-k-1)}{2(k+1)(k+2)(2k+3)} [ M^2-(k+1)^2
\lambda^2 ] \nonumber
\\
&a_0{}^2 = \dfrac{(s+1)(s-1)}{3} [ M^2- \lambda^2 ]
\\
& b_k = \dfrac{s^2M^2}{4k(k+1)^2},\qquad b_0=\dfrac{sMa_0}{2},
\qquad M^2 = m^2 + (s-1)^2 \lambda^2\nonumber
\end{eqnarray}
It is invariant under the following gauge transformations
\begin{eqnarray}
\delta \Omega^{\alpha(2k)} &=& D \eta^{\alpha(2k)} +
\frac{(k+2)a_k}{k} e_{\beta(2)} \eta^{\alpha(2k)\beta(2)} \nonumber \\
 && + \frac{a_{k-1}}{k(2k-1)} e^{\alpha(2)} \eta^{\alpha(2k-2)}
 + \frac{b_k}{k} e^\alpha{}_\beta \xi^{\alpha(2k-1)\beta} \nonumber \\
\delta f^{\alpha(2k)} &=& D \xi^{\alpha(2k)} + e^\alpha{}_\beta
\eta^{\alpha(2k-1)\beta} + a_k e_{\beta(2)}
\xi^{\alpha(2k)\beta(2)} \nonumber \\
 && + \frac{(k+1)a_{k-1}}{k(k-1)(2k-1)} e^{\alpha(2)}
\xi^{\alpha(2k-2)} \nonumber \\
\delta \Omega^{\alpha(2)} &=& D \eta^{\alpha(2)} + 3a_1 e_{\beta(2)}
\eta^{\alpha(2)\beta(2)} + b_1
e^\alpha{}_\gamma \xi^{\alpha\gamma} \\
\delta f^{\alpha(2)} &=& D \xi^{\alpha(2)} + e^\alpha{}_\gamma
\eta^{\alpha\gamma} + a_1 e_{\beta(2)}
\xi^{\alpha(2)\beta(2)}  + 2a_0 e^{\alpha(2)} \xi \nonumber \\
\delta B^{\alpha(2)} &=& 2a_0 \eta^{\alpha(2)}, \qquad \delta A = D
\xi + \frac{a_0}{4} e_{\alpha(2)} \xi^{\alpha(2)}
\nonumber \\
\delta \pi^{\alpha(2)} &=& \frac{Msa_0}{2} \xi^{\alpha(2)}, \qquad
\delta \varphi = - 2Ms \xi \nonumber
\end{eqnarray}
One can construct the curvatures invariant under these gauge
transformation. After change of normalization
\begin{eqnarray}\label{Norm_s}
B^{\alpha(2)}\rightarrow 2a_0B^{\alpha(2r)}\qquad
\pi^{\alpha(2)}\rightarrow b_0\pi^{\alpha(2)}
\end{eqnarray}
the curvatures look like
\begin{eqnarray}\label{Curv_s1}
{\cal{R}}^{\alpha(2k)} &=& D \Omega^{\alpha(2k)} +
\frac{(k+2)a_k}{k} e_{\beta(2)} \Omega^{\alpha(2k)\beta(2)} \nonumber
\\
 && + \frac{a_{k-1}}{k(2k-1)} e^{\alpha(2)} \Omega^{\alpha(2k-2)}
 + \frac{b_k}{k} e^\alpha{}_\beta f^{\alpha(2k-1)\beta} \nonumber \\
{\cal{T}}^{\alpha(2k)} &=& D f^{\alpha(2k)} + e^\alpha{}_\beta
\Omega^{\alpha(2k-1)\beta} + a_k e_{\beta(2)}
f^{\alpha(2k)\beta(2)} \nonumber \\
 && + \frac{(k+1)a_{k-1}}{k(k-1)(2k-1)} e^{\alpha(2)}
f^{\alpha(2k-2)} \nonumber \\
{\cal{R}}^{\alpha(2)} &=& D \Omega^{\alpha(2)} + 3a_1 e_{\beta(2)}
\Omega^{\alpha(2)\beta(2)} + b_1 e^\alpha{}_\gamma
f^{\alpha\gamma}-{a_0}^2E^\alpha{}_\beta B^{\alpha\beta}
+b_0E^{\alpha(2)}\varphi \\
{\cal{T}}^{\alpha(2)} &=& Df^{\alpha(2)} + e^\alpha{}_\gamma
\Omega^{\alpha\gamma} + a_1 e_{\beta(2)}
f^{\alpha(2)\beta(2)}  + 2a_0 e^{\alpha(2)}A \nonumber \\
{\cal{B}}^{\alpha(2)}
&=&DB^{\alpha(2)}-\Omega^{\alpha(2)}+{b_1}e^\alpha{}_\beta\pi^{\alpha\beta}+
3a_1e_{\beta(2)}B^{\alpha(2)\beta(2)} \nonumber
\\
{\cal{A}} &=& DA + \frac{a_0}{4} e_{\alpha(2)}
f^{\alpha(2)}-2a_0E_{\gamma(2)}B^{\gamma(2)}
\nonumber\\
\Pi^{\alpha(2)} &=&D\pi^{\alpha(2)}-f^{\alpha(2)}+e^\alpha{}_\beta
B^{\alpha\beta}-\frac{a_0}{sM}e^{\alpha(2)}\varphi+a_1e_{\beta(2)}\pi^{\alpha(2)\beta(2)}
\nonumber\\
\Phi &=&D\varphi+ 2MsA-b_0e_{\alpha(2)}\pi^{\alpha(2)} \nonumber
\end{eqnarray}
\begin{eqnarray}\label{Curv_s2}
{\cal{B}}^{\alpha(2k)}
&=&DB^{\alpha(2k)}-\Omega^{\alpha(2k)}+\frac{b_k}{k}e^\alpha{}_\beta\pi^{\alpha(2k-1)\beta}+
\frac{a_{k-1}}{k(2k-1)}e^{\alpha(2)}B^{\alpha(2k-2)}\nonumber\\
&&+\frac{(k+2)}{k}a_ke_{\beta(2)}B^{\alpha(2k)\beta(2)} \nonumber
\nonumber\\
\Pi^{\alpha(2k)}
&=&D\pi^{\alpha(2k)}-f^{\alpha(2k)}+e^\alpha{}_\beta
B^{\alpha(2k-1)\beta}+\frac{(k+1)a_{k-1}}{k(k-1)(2k-1)}e^{\alpha(2)}\pi^{\alpha(2k-2)}\nonumber\\
&&+a_ke_{\beta(2)}\pi^{\alpha(2k)\beta(2)}
\end{eqnarray}
In comparison with massive spin 2 case, the construction of
curvatures for higher spins has some peculiarities. Namely, in order
to achieve gauge invariance for all curvatures we should introduce
the so called extra fields $B^{\alpha(2k)},\pi^{\alpha(2k)}$, $2\leq
k\leq s-1$ with the following gauge transformations
$$
\delta B^{\alpha(2k)}=\eta^{\alpha(2k)}\qquad \delta
\pi^{\alpha(2k)}=\xi^{\alpha(2k)}
$$
As we already pointed out above, in three dimensions it is possible
to write Lagrangian in terms of the curvatures only. In case of
arbitrary integer spin field, the  corresponding Lagrangian
(\ref{Lag_s}) can be rewritten in the form
\begin{eqnarray}\label{LagC_s}
{\cal{L}}&=&-\frac{1}{2}\sum_{k=1}^{s-1}(-1)^{k+1}[{\cal{R}}_{\alpha(2k)}\Pi^{\alpha(2k)}
+{\cal{T}}_{\alpha(2k)}{\cal{B}}^{\alpha(2k)}]
+\frac{a_0}{2sM}e_{\alpha(2)}{\cal{B}}^{\alpha(2)}\Phi
\end{eqnarray}

\subsection{Fermionic spin-$(s+\frac12)$ field}

In gauge invariant form the massive spin $(s+1/2)$ field is
described by system of massless fields with spins
$(s+1/2),(s-1/2),...,1/2$. In frame-like approach the corresponding
set of fields consists of (\ref{massless_fermions})
$$
\Phi^{\alpha(2k+1)}\quad0\leq k\leq (s-1),\quad \phi^\alpha
$$
Free Lagrangian for fields with mass $m_1$ in $AdS_3$ looks like
\begin{eqnarray}\label{Lag_s/2}
\frac{1}{i} {\cal L} &=& \sum_{k=0}^{s-1} (-1)^{k+1} [ \frac{1}{2}
\Phi_{\alpha(2k+1)} D \Phi^{\alpha(2k+1)} ]
+ \frac{1}{2} \phi_\alpha E^\alpha{}_\beta D \phi^\beta \nonumber \\
 && + \sum_{k=1}^{s-1} (-1)^{k+1} c_k \Phi_{\alpha(2k-1)\beta(2)}
e^{\beta(2)} \Phi^{\alpha(2k-1)} + c_0 \Phi_\alpha E^\alpha{}_\beta
\phi^\beta \nonumber \\
 && + \sum_{k=0}^{s-1} (-1)^{k+1} \frac{d_k}{2}
\Phi_{\alpha(2k)\beta} e^\beta{}_\gamma \Phi^{\alpha(2k)\gamma} -
\frac{3d_0}{2} E \phi_\alpha \phi^\alpha
\end{eqnarray}
where
\begin{eqnarray}\label{fermion_data}
&c_k{}^2 = \dfrac{(s+k+1)(s-k)}{2(k+1)(2k+1)} [ M_1{}^2 - (2k+1)^2
\frac{\lambda^2}{4} ]
\nonumber\\
& c_0{}^2 = 2s(s+1) [ M_1{}^2 - \frac{\lambda^2}{4} ]
\\
& d_k = \dfrac{(2s+1)}{(2k+3)} M_1, \qquad M_1{}^2 = m_1{}^2 +
(s-\frac{1}{2})^2 \lambda^2\nonumber
\end{eqnarray}
The Lagrangian is invariant under the gauge transformations
\begin{eqnarray}\label{GT_s/2}
\delta \Phi^{\alpha(2k+1)} &=& D \xi^{\alpha(2k+1)} +
\frac{d_k}{(2k+1)} e^\alpha{}_\beta \xi^{\alpha(2k)\beta} \nonumber \\
 && + \frac{c_k}{k(2k+1)} e^{\alpha(2)} \xi^{\alpha(2k-1)}
 + c_{k+1} e_{\beta(2)} \xi^{\alpha(2k+1)\beta(2)} \\
\delta \phi^\alpha &=& c_0 \xi^\alpha \nonumber
\end{eqnarray}
Let us construct gauge invariant curvatures with respect
(\ref{GT_s/2}). After change of normalization
$\phi^{\alpha}\rightarrow c_0\phi^{\alpha}$ the curvatures have form
\begin{eqnarray}\label{Curv_s/2}
{\cal{F}}^{\alpha(2k+1)} &=& D \Phi^{\alpha(2k+1)} +
\frac{d_k}{(2k+1)} e^\alpha{}_\beta \Phi^{\alpha(2k)\beta} \nonumber
\\
 && + \frac{c_k}{k(2k+1)} e^{\alpha(2)} \Phi^{\alpha(2k-1)}
 + c_{k+1} e_{\beta(2)} \Phi^{\alpha(2k+1)\beta(2)} \nonumber\\
{\cal{F}}^{\alpha} &=& D \Phi^{\alpha} + {d_0} e^\alpha{}_\beta
\Phi^{\beta} + c_{1} e_{\beta(2)} \Phi^{\alpha\beta(2)}
-c_0{}^2E^\alpha{}_\beta\phi^\beta
\nonumber\\
{\cal{C}}^\alpha&=&D\phi^\alpha-\Phi^\alpha+d_0e^\alpha{}_\beta\phi^\beta
+c_1e_{\beta(2)}\phi^{\alpha\beta(2)}
\nonumber\\
{\cal{C}}^{\alpha(2k+1)}&=&D\phi^{\alpha(2k+1)}-\Phi^{\alpha(2k+1)}
+\frac{d_k}{(2k+1)}e^\alpha{}_\beta\phi^{\alpha(2k)\beta}\nonumber\\
&& +\frac{c_k}{k(2k+1)}e^{\alpha(2)}\phi^{\alpha(2k-1)}
+c_{k+1}e_{\beta(2)}\phi^{\alpha(2k+1)\beta(2)}
\end{eqnarray}
As in the case of integer spins in order to achieve gauge invariance
for all curvatures we should introduce the set of extra fields
$\phi^{\alpha(2k+1)}$, $1\leq k\leq (s-1)$ with the following gauge
transformation
$$
\delta\phi^{\alpha(2k+1)}=\xi^{\alpha(2k+1)}
$$
Finally the Lagrangian (\ref{Lag_s/2}) can be rewritten in terms of
curvatures only as follows
\begin{eqnarray}\label{LagC_s/2}
{\cal{L}}&=&-\frac{i}{2}\sum_{k=0}^{s-2}(-1)^{k+1}{\cal{F}}_{\alpha(2k+1)}{\cal{C}}^{\alpha(2k+1)}
\end{eqnarray}
In section 5 we will need the description of massive spin-$(s-1/2)$
field. It can be obtained from the above description by replacement
$s\rightarrow(s-1)$.

So we have considered the free massive fields with spins $s$ and
spins $s+\frac12$. Also, we have formulated the gauge invariant
curvatures (\ref{Curv_s1}), (\ref{Curv_s2}), (\ref{Curv_s/2}) and
gauge invariant Lagrangians (\ref{LagC_s}), (\ref{LagC_s/2}). In the
next sections we  will study a supersymmetrization of these results
and construct the Lagrangian description of the supermultiplets
$(s,s+\frac12)$ $(s,s-\frac12)$.

\section{Massive supermultiplet $(s,s+\frac12)$}

In this section we consider the massive higher spin supermultiplets
when the highest spin is fermion. For these supermultiplets we
construct the deformation of the curvatures, find the
supertransformations and present the supersymmetric Lagrangian.

{\bf Massless flat limit}
\\
Before we turn to realization of given massive supermultiplets let
us consider their structure at massless flat limit
$m,m_1,\lambda\rightarrow0$. In this case the Lagrangian will be
described by the system of massless fields with spins
$(s+\frac12),s,...,\frac12,0$ in three dimensional flat space
\begin{eqnarray}
{\cal L} &=& \sum_{k=1}^{s-1} (-1)^{k+1} [ k
\Omega_{\alpha(2k-1)\beta} e^\beta{}_\gamma
\Omega^{\alpha(2k-1)\gamma} + \Omega_{\alpha(2k)} D f^{\alpha(2k)}
 ] \nonumber \\
 && + E B_{\alpha\beta} B^{\alpha\beta} - B_{\alpha\beta}
e^{\alpha\beta} D A - E \pi_{\alpha\beta} \pi^{\alpha\beta} +
\pi_{\alpha\beta} E^{\alpha\beta} D \varphi \nonumber \\
&&+\frac{i}{2}\sum_{k=0}^{s-1} (-1)^{k+1} \Phi_{\alpha(2k+1)} D
\Phi^{\alpha(2k+1)}  + \frac{1}{2} \phi_\alpha E^\alpha{}_\beta D
\phi^\beta
\end{eqnarray}
One can show that this Lagrangian is supersymmetric. Indeed, if the
equations of motion (\ref{Low_eq}) are fulfilled, the Lagrangian is
invariant under the following supertransformations
\begin{eqnarray*}
\delta f^{\alpha(2k)} &=& i\beta_k\Phi^{\alpha(2k-1)}\zeta^\alpha
+i\alpha_k\Phi^{\alpha(2k)\beta}\zeta_\beta
\\
\delta f^{\alpha(2)} &=&i\beta_1\Phi^{\alpha}\zeta^\alpha
+i\alpha_1\Phi^{\alpha(2)\beta}\zeta_\beta
\\
\delta A &=&i\alpha_0\Phi^\alpha\zeta_\alpha
+ic_0\beta_0e_{\alpha\beta}\phi^\alpha\zeta^\beta ,\qquad
\delta\varphi =-\frac{ic_0\tilde\delta_0}{2}\phi^\gamma\zeta_\gamma
\\
\delta\Phi^{\alpha(2k+1)} &=&
\frac{\alpha_k}{(2k+1)}\Omega^{\alpha(2k)}\zeta^\alpha+
2(k+1)\beta_{k+1}\Omega^{\alpha(2k+1)\beta}\zeta_\beta
\\
\delta\Phi^{\alpha} &=&
2\beta_1\Omega^{\alpha\beta}\zeta_\beta+2a_0\alpha_0e_{\beta(2)}B^{\beta(2)}\zeta^\alpha
\\
\delta\phi^\alpha&=&\frac{8a_0\beta_0}{c_0}B^{\alpha\beta}\zeta_\beta
+\frac{b_0\ti\delta_0}{c_0}\pi^{\alpha\beta}\zeta_\beta
\end{eqnarray*}
Here we take into account the normalization (\ref{Norm_s}). Thus,
requiring that massive theory has a correct massless flat limit we
partially fix an arbitrariness in the choice of the
supertransformations. Parameters
$\alpha_k,\beta_k,\beta_0,\alpha_0,\ti\delta_0$ at this step are still
arbitrary.
\\
{\bf Deformation of curvatures}
\\
Again we will realize supersymmetry deforming the
curvatures by the background gravitino field $\Psi^\alpha$. We start
with the construction of the deformations for bosonic fields
\begin{eqnarray*}
\Delta{\cal{R}}^{\alpha(2k)} &=&
i\rho_k\Phi^{\alpha(2k-1)}\Psi^\alpha+i\sigma_k\Phi^{\alpha(2k)\beta}\Psi_\beta
\\
\Delta{\cal{T}}^{\alpha(2k)} &=&
i\beta_k\Phi^{\alpha(2k-1)}\Psi^\alpha
+i\alpha_k\Phi^{\alpha(2k)\beta}\Psi_\beta
\\
\Delta{\cal{R}}^{\alpha(2)} &=& i\rho_1\Phi^{\alpha}\Psi^\alpha
+i\sigma_1\Phi^{\alpha(2)\beta}\Psi_\beta+i\hat\rho_0e^{\alpha(2)}\phi^\beta\Psi_\beta \\
\Delta{\cal{T}}^{\alpha(2)} &=&i\beta_1\Phi^{\alpha}\Psi^\alpha
+i\alpha_1\Phi^{\alpha(2)\beta}\Psi_\beta
\\
\Delta{\cal{A}} &=&i\alpha_0\Phi^\alpha\Psi_\alpha
+ic_0\beta_0e_{\alpha(2)}\phi^\alpha\Psi^\beta ,\qquad \Delta\Phi
=\frac{ic_0\tilde\delta_0}{2}\phi^\alpha\Psi_\alpha
\\
\Delta{\cal{B}}^{\alpha(2k)}
&=&-i\hat\rho_k\phi^{\alpha(2k-1)}\Psi^\alpha-i\hat\sigma_k\phi^{\alpha(2k)\beta}\Psi_\beta
\\
\Delta\Pi^{\alpha(2k)} &=&
-i\hat\beta_k\phi^{\alpha(2k-1)}\Psi^\alpha-i\hat\alpha_k\phi^{\alpha(2k)\beta}\Psi_\beta
\end{eqnarray*}
Corresponding ansatz for supertransformations has the form
\begin{eqnarray}\label{Super_b}
\delta\Omega^{\alpha(2k)} &=&
i\rho_k\Phi^{\alpha(2k-1)}\zeta^\alpha+i\sigma_k\Phi^{\alpha(2k)\beta}\zeta_\beta
\nonumber\\
\delta f^{\alpha(2k)} &=& i\beta_k\Phi^{\alpha(2k-1)}\zeta^\alpha
+i\alpha_k\Phi^{\alpha(2k)\beta}\zeta_\beta
\nonumber\\
\delta\Omega^{\alpha(2)} &=& i\rho_1\Phi^{\alpha}\zeta^\alpha
+i\sigma_1\Phi^{\alpha(2)\beta}\zeta_\beta+i\hat\rho_0e^{\alpha(2)}\phi^\beta\zeta_\beta
\nonumber \\
\delta f^{\alpha(2)} &=&i\beta_1\Phi^{\alpha}\zeta^\alpha
+i\alpha_1\Phi^{\alpha(2)\beta}\zeta_\beta
\\
\delta A &=&i\alpha_0\Phi^\alpha\zeta_\alpha
+ic_0\beta_0e_{\alpha(2)}\phi^\alpha\zeta^\beta ,\qquad
\delta\varphi =-\frac{ic_0\tilde\delta_0}{2}\phi^\gamma\zeta_\gamma
\nonumber\\
\delta B^{\alpha(2k)}
&=&i\hat\rho_k\phi^{\alpha(2k-1)}\zeta^\alpha+i\hat\sigma_k\phi^{\alpha(2k)\beta}\zeta_\beta
\nonumber\\
\delta\pi^{\alpha(2k)} &=&
i\hat\beta_k\phi^{\alpha(2k-1)}\zeta^\alpha+i\hat\alpha_k\phi^{\alpha(2k)\beta}\zeta_\beta
\nonumber
\end{eqnarray}
All parameters will be fixed from requirement of covariant
transformations of the curvatures (\ref{CReq}). First of all we
consider
\begin{eqnarray}\label{CT_HS1}
\delta\hat{\cal{R}}^{\alpha(2k)} &=&
i\rho_k{\cal{F}}^{\alpha(2k-1)}\zeta^\alpha+i\sigma_k{\cal{F}}^{\alpha(2k)\beta}\zeta_\beta
\nonumber\\
\delta\hat{\cal{T}}^{\alpha(2k)} &=&
i\beta_k{\cal{F}}^{\alpha(2k-1)}\zeta^\alpha+i\alpha_k{\cal{F}}^{\alpha(2k)\beta}\zeta_\beta
\end{eqnarray}
It leads to relation $M_1=M+\frac{\lambda}{2}$ between mass
parameters $M_1$ and $M$ and defines the parameters
\begin{eqnarray}\label{solution1}
\alpha_{k}{}^2&=&k(s+k+1) [ M+(k+1) \lambda ]\hat\alpha^2
\nonumber\\
\beta_{k}{}^2&=& \frac{(k+1)(s-k)}{k(2k+1)} [ M-k\lambda
]\hat\beta^2
\nonumber\\
\sigma_{k}{}^2&=&\frac{(s+k+1)}{k(k+1)^2} [ M+(k+1) \lambda
]\hat\sigma^2
\nonumber\\
\rho_{k}{}^2&=&\frac{(s-k)}{k^3(k+1)(2k+1)} [ M-k\lambda]\hat\rho^2
\end{eqnarray}
where
$$
\hat\beta=\frac{\hat\alpha}{\sqrt2},\qquad
\hat\rho=\frac{sM}{2\sqrt2}\hat\alpha,\qquad
\hat\sigma=\frac{sM}{2}\hat\alpha,\qquad
\hat\alpha^2=\frac{\alpha_{s-1}{}^2}{2s(s-1)[ M+s \lambda ]}
$$
From the requirement that
\begin{eqnarray}\label{CT_HS2}
\delta\hat{\cal{B}}^{\alpha(2k)}&=&i\hat\rho_k{\cal{C}}^{\alpha(2k-1)}\zeta^\alpha
+i\hat\sigma_k{\cal{C}}^{\alpha(2k)\beta}\zeta_\beta
\nonumber\\
\delta\hat\Pi^{\alpha(2k)}&=&i\hat\beta_k{\cal{C}}^{\alpha(2k-1)}\zeta^\alpha
+i\hat\alpha_k{\cal{C}}^{\alpha(2k)\beta}\zeta_\beta
\end{eqnarray}
one gets
$$
\hat\rho_k=\rho_k,\qquad
\hat\sigma_k=\sigma_k,\qquad\hat\beta_k=\beta_k,\qquad
\hat\alpha_k=\alpha_k
$$
Requirement of covariant transformations for the other curvatures
\begin{eqnarray}\label{CT_HS3}
\delta\hat{\cal{R}}^{\alpha(2)} &=&
i\rho_1{\cal{F}}^{\alpha}\zeta^\alpha
+i\sigma_1{\cal{F}}^{\alpha(2)\beta}\zeta_\beta-i\hat\rho_0e^{\alpha(2)}{\cal{C}}^\beta\zeta_\beta
\nonumber\\
\delta\hat{\cal{T}}^{\alpha(2)} &=&
i\beta_1{\cal{F}}^{\alpha}\zeta^\alpha
+i\alpha_1{\cal{F}}^{\alpha(2)\beta}\zeta_\beta\nonumber\\
\delta\hat{\cal{A}} &=& i\alpha_0{\cal{F}}^\alpha\zeta_\alpha
-ic_0\beta_0e_{\alpha\beta}{\cal{C}}^\alpha\zeta^\beta,\qquad
\delta\hat\Phi
=-\frac{ic_0\tilde\delta_0}{2}{\cal{C}}^\gamma\zeta_\gamma
\end{eqnarray}
gives the solution
$$
\h\rho_0=-\frac18c_0{}^2\beta_1,\qquad\ti\delta_0=4\beta_0=\frac{c_0}{a_0}\beta_1,
\qquad\alpha_0=\frac{c_0{}^2}{4sMa_0}\beta_1
$$

Now let us consider the deformation of curvatures for fermion. We
choose ansatz in the form
\begin{eqnarray*}
\Delta{\cal{F}}^{\alpha(2k+1)} &=&
\frac{\alpha_k}{(2k+1)}\Omega^{\alpha(2k)}\Psi^\alpha+
2(k+1)\beta_{k+1}\Omega^{\alpha(2k+1)\beta}\Psi_\beta\\
&& +\gamma_kf^{\alpha(2k)}\Psi^\alpha+
\delta_kf^{\alpha(2k+1)\beta}\Psi_\beta
\\
\Delta{\cal{F}}^{\alpha} &=&
2\beta_{1}\Omega^{\alpha\beta}\Psi_\beta+2a_0\alpha_0e_{\beta(2)}B^{\beta(2)}\Psi^\alpha
+ \delta_0f^{\alpha\beta}\Psi_\beta+\gamma_0A\Psi^\alpha+
\tilde\gamma_0\varphi e^\alpha{}_\beta\Psi^\beta
\\
\Delta{\cal{C}}^\alpha&=&-\frac{8a_0\beta_0}{c_0}B^{\alpha\beta}\Psi_\beta
-\frac{b_0\tilde\delta_0}{c_0}\pi^{\alpha\beta}\Psi_\beta
-\rho_0\varphi\Psi^\alpha
\\
\Delta{\cal{C}}^{\alpha(2k+1)}&=&-\tilde\beta_kB^{\alpha(2k+1)\beta}\Psi_\beta
-\tilde\alpha_k
B^{\alpha(2k)}\Psi^\alpha-\tilde\delta_k\pi^{\alpha(2k+1)\beta}\Psi_\beta
-\tilde\gamma_k\pi^{\alpha(2k)}\Psi^\alpha
\end{eqnarray*}
\\
and ansatz for supertransformations in the form
\begin{eqnarray}\label{Super_f}
\delta\Phi^{\alpha(2k+1)} &=&
\frac{\alpha_k}{(2k+1)}\Omega^{\alpha(2k)}\zeta^\alpha+
2(k+1)\beta_{k+1}\Omega^{\alpha(2k+1)\beta}\zeta_\beta\nonumber\\
&& +\gamma_kf^{\alpha(2k)}\zeta^\alpha+
\delta_kf^{\alpha(2k+1)\beta}\zeta_\beta
\nonumber\\
\delta\Phi^{\alpha} &=&
2\beta_{1}\Omega^{\alpha\beta}\zeta_\beta+2a_0\alpha_0e_{\beta(2)}B^{\beta(2)}\zeta^\alpha
+ \delta_0f^{\alpha\beta}\zeta_\beta+\gamma_0A\zeta^\alpha+
\tilde\gamma_0\varphi e^\alpha{}_\beta\zeta^\beta
\\
\delta\phi^\alpha&=&\frac{8a_0\beta_0}{c_0}B^{\alpha\beta}\zeta_\beta+
\frac{b_0\tilde\delta_0}{c_0}\pi^{\alpha\beta}\zeta_\beta
+\rho_0\varphi\zeta^\alpha
\nonumber\\
\delta\phi^{\alpha(2k+1)}&=&\tilde\beta_kB^{\alpha(2k+1)\beta}\zeta_\beta
+\tilde\alpha_k
B^{\alpha(2k)}\zeta^\alpha+\tilde\delta_k\pi^{\alpha(2k+1)\beta}\zeta_\beta
+\tilde\gamma_k\pi^{\alpha(2k)}\zeta^\alpha\nonumber
\end{eqnarray}

From the requirement that
\begin{eqnarray}\label{CT_HS4}
\delta\hat{\cal{F}}^{\alpha(2k+1)} &=&
\frac{\alpha_k}{(2k+1)}{\cal{R}}^{\alpha(2k)}\zeta^\alpha+
2(k+1)\beta_{k+1}{\cal{R}}^{\alpha(2k+1)\beta}\zeta_\beta\nonumber\\
&& +\gamma_k{\cal{T}}^{\alpha(2k)}\zeta^\alpha+
\delta_k{\cal{T}}^{\alpha(2k+1)\beta}\zeta_\beta
\end{eqnarray}
we have the same relation between masses $M_1=M+\frac{\lambda}{2}$.
Besides, it leads to
\begin{eqnarray*}
\gamma_{k}{}^2&=& \frac{(s+k+1)}{k(k+1)^2(2k+1)^2}[ M+(k+1) \lambda
]\hat\gamma^2
\\
\delta_{k}{}^2&=&\frac{(s-k-1)}{(k+1)(k+2)(2k+3)}[
M-(k+1)\lambda]\hat\delta^2
\end{eqnarray*}
where
$$
\hat\gamma=\frac{sM}{2}\hat\alpha,\qquad\hat\delta=\frac{sM}{\sqrt2}\hat\alpha
$$
Requirement
\begin{eqnarray}\label{CT_HS5}
\delta\hat{\cal{C}}^{\alpha(2k+1)}&=&\tilde\beta_k{\cal{B}}^{\alpha(2k+1)\beta}\zeta_\beta
+\tilde\alpha_k
{\cal{B}}^{\alpha(2k)}\zeta^\alpha+\tilde\delta_k\Pi^{\alpha(2k+1)\beta}\zeta_\beta
+\tilde\gamma_k\Pi^{\alpha(2k)}\zeta^\alpha
\end{eqnarray}
gives us
$$
\tilde\gamma_k=\gamma_k,\quad\tilde\delta_k=\delta_k,\quad
\tilde\alpha_k=\frac{\alpha_k}{(2k+1)},\quad\tilde\beta_k=2(k+1)\beta_{k+1}
$$
At last, requirement for the other curvatures
\begin{eqnarray}\label{CT_HS6}
\delta\hat{\cal{F}}^{\alpha}
&=&2\beta_{1}{\cal{R}}^{\alpha\beta}\zeta_\beta-2a_0\alpha_0e_{\beta(2)}{\cal{B}}^{\beta(2)}\zeta^\alpha
+
\delta_0{\cal{T}}^{\alpha\beta}\zeta_\beta+\gamma_0{\cal{A}}\zeta^\alpha+
\tilde\gamma_0\Phi e^\alpha{}_\beta\zeta^\beta
\nonumber\\
\delta\hat{\cal{C}}^\alpha&=&\frac{8a_0\beta_0}{c_0}{\cal{B}}^{\alpha\beta}\zeta_\beta
+\frac{b_0\ti\delta_0}{c_0}\Pi^{\alpha\beta}\zeta_\beta
+\rho_0\Phi\zeta^\alpha
\end{eqnarray}
yields solution
$$
\gamma_0=-2\ti\gamma_0=\frac{c_0{}^2}{2a_0}\beta_1,\quad
\rho_0=-\frac{c_0{}^2}{4sMa_0}\beta_1
$$
Now, all the arbitrary parameters are fixed.

Supersymmetric Lagrangian is the sum of free Lagrangians where the
initial curvatures are replacement by deformed ones
\begin{eqnarray}\label{LagSC_s}
\h{\cal{L}}&=&-\frac{1}{2}\sum_{k=1}^{s-1}(-1)^{k+1}[\h{\cal{R}}_{\alpha(2k)}\h\Pi^{\alpha(2k)}
+\h{\cal{T}}_{\alpha(2k)}\h{\cal{B}}^{\alpha(2k)}]
+\frac{a_0}{2sM}e_{\alpha(2)}\h{\cal{B}}^{\alpha(2)}\h\Phi\nonumber\\
&&-\frac{i}{2}\sum_{k=0}^{s-1}(-1)^{k+1}\h{\cal{F}}_{\alpha(2k+1)}\h{\cal{C}}^{\alpha(2k+1)}
\end{eqnarray}
The Lagrangian is invariant under the supertransformations
(\ref{CT_HS1}),(\ref{CT_HS2}),(\ref{CT_HS3}),(\ref{CT_HS4}),(\ref{CT_HS5}),(\ref{CT_HS6})
up to equations of motion for the fields
$B^{\alpha(2)},\pi^{\alpha(2)}$
\begin{eqnarray}\label{Eq_low}
\Phi=0,\quad{\cal{A}}=0\quad\Rightarrow\quad
e_{\gamma(2)}\Pi^{\gamma(2)}=D\Phi-2sM{\cal{A}}=0
\end{eqnarray}

The Lagrangian (\ref{LagSC_s}) is a final solution for the massive
supermultiplet $(s,s+\frac12)$.

\section{Massive supermultiplet $(s,s-\frac12)$}

In this section we consider another massive higher spin
supermultiplet when the highest spin is boson. The massive spin-$s$
field was described in section 3.1 in terms of massless fields. The
massive spin-$(s-1/2)$ field can be obtained for the results in
section 3.2 if one makes the replacement $s\rightarrow(s-1)$. So the
set of massless fields for the massive field with spin $s-1/2$ is
$\Phi^{\alpha(2k+1)}$, $0 \le k \le s-2$ and $\phi^{\alpha}$. The
gauge invariant curvatures and the Lagrangian have the forms
(\ref{GT_s/2}) and (\ref{LagC_s/2}) where the parameters are
\begin{eqnarray}\label{InD}
&c_k{}^2 = \dfrac{(s+k)(s-k-1)}{2(k+1)(2k+1)} [ M_1{}^2 - (2k+1)^2
\frac{\lambda^2}{4} ]
\nonumber\\
&c_0{}^2 = 2s(s-1) [ M_1{}^2 - \frac{\lambda^2}{4} ]
\\
&d_k = \dfrac{(2s-1)}{(2k+3)} M_1, \qquad M_1{}^2 = m_1{}^2 +
(s-\frac{3}{2})^2 \lambda^2\nonumber
\end{eqnarray}

Following our procedure we should construct the supersymmetric
deformations for curvatures. Actually the structure of deformed
curvatures and supertransformations have the same form as in
previous section for supermultiplets $(s,s+1/2)$. There is a
difference only in parameters (\ref{InD}). Therefore we present here
only the supertransformations for curvatures. Requirement
(\ref{CReq}) for bosonic fields
\begin{eqnarray*}
\delta\hat{\cal{R}}^{\alpha(2k)} &=&
i\rho_k{\cal{F}}^{\alpha(2k-1)}\zeta^\alpha+i\sigma_k{\cal{F}}^{\alpha(2k)\beta}\zeta_\beta
\\
\delta\hat{\cal{T}}^{\alpha(2k)} &=&
i\beta_k{\cal{F}}^{\alpha(2k-1)}\zeta^\alpha+i\alpha_k{\cal{F}}^{\alpha(2k)\beta}\zeta_\beta
\\
\delta\hat{\cal{R}}^{\alpha(2)} &=&
i\rho_1{\cal{F}}^{\alpha}\zeta^\alpha
+i\sigma_1{\cal{F}}^{\alpha(2)\beta}\zeta_\beta-i\hat\rho_0e^{\alpha(2)}{\cal{C}}^\beta\zeta_\beta
\\
\delta\hat{\cal{T}}^{\alpha(2)} &=&
i\beta_1{\cal{F}}^{\alpha}\zeta^\alpha
+i\alpha_1{\cal{F}}^{\alpha(2)\beta}\zeta_\beta\\
\delta\hat{\cal{A}} &=& i\alpha_0{\cal{F}}^\alpha\zeta_\alpha
-ic_0\beta_0e_{\alpha\beta}{\cal{C}}^\alpha\zeta^\beta,\qquad
\delta\hat\Phi
=-\frac{ic_0\tilde\delta_0}{2}{\cal{C}}^\gamma\zeta_\gamma
\\
\delta\hat{\cal{B}}^{\alpha(2k)}&=&i\hat\rho_k{\cal{C}}^{\alpha(2k-1)}\zeta^\alpha
+i\hat\sigma_k{\cal{C}}^{\alpha(2k)\beta}\zeta_\beta
\\
\delta\hat\Pi^{\alpha(2k)}&=&i\hat\beta_k{\cal{C}}^{\alpha(2k-1)}\zeta^\alpha
+i\hat\alpha_k{\cal{C}}^{\alpha(2k)\beta}\zeta_\beta
\end{eqnarray*}
gives us the relation $M_1=M-\frac{\lambda}{2}$ between masses $M_1$
and $M$.  Besides, it leads to
\begin{eqnarray}\label{SupMult2}
\sigma_{k}{}^2&=&\frac{(s-k-1)}{k(k+1)^2} [ M-(k+1) \lambda
]\hat\sigma^2\nonumber \\
\rho_{k}{}^2&=&\frac{(s+k)}{k^3(k+1)(2k+1)} [ M+k\lambda]\hat\rho^2
\nonumber\\
\alpha_{k}{}^2&=&k(s-k-1) [ M-(k+1) \lambda ]\hat\alpha^2
\nonumber\\
\beta_{k}{}^2&=& \frac{(k+1)(s+k)}{k(2k+1)} [ M+k\lambda
]\hat\beta^2
\end{eqnarray}
$$
\hat\rho_k=\rho_k,\qquad
\hat\sigma_k=\sigma_k,\qquad\hat\beta_k=\beta_k,\qquad
\hat\alpha_k=\alpha_k
$$
$$
\h\rho_0=-\frac18c_0{}^2\beta_1,\qquad\ti\delta_0=4\beta_0=\frac{c_0}{a_0}\beta_1,
\qquad\alpha_0=\frac{c_0{}^2}{4sMa_0}\beta_1
$$
where
$$
\hat\beta=\frac{\hat\alpha}{\sqrt2},\qquad
\hat\rho=\frac{sM}{2\sqrt2}\hat\alpha,\qquad
\hat\sigma=\frac{sM}{2}\hat\alpha,\qquad
\hat\alpha^2=\frac{\alpha_{s-2}{}^2}{(s-2)[ M-(s-1) \lambda ]}
$$
From requirement of covariant supertransformations for fermionic
curvatures
\begin{eqnarray*}
\delta\hat{\cal{F}}^{\alpha(2k+1)} &=&
\frac{\alpha_k}{(2k+1)}{\cal{R}}^{\alpha(2k)}\zeta^\alpha+
2(k+1)\beta_{k+1}{\cal{R}}^{\alpha(2k+1)\beta}\zeta_\beta\\
&& +\gamma_k{\cal{T}}^{\alpha(2k)}\zeta^\alpha+
\delta_k{\cal{T}}^{\alpha(2k+1)\beta}\zeta_\beta
\\
\delta\hat{\cal{F}}^{\alpha}
&=&2\beta_{1}{\cal{R}}^{\alpha\beta}\zeta_\beta-2a_0\alpha_0e_{\beta(2)}{\cal{B}}^{\beta(2)}\zeta^\alpha
+
\delta_0{\cal{T}}^{\alpha\beta}\zeta_\beta+\gamma_0{\cal{A}}\zeta^\alpha+
\tilde\gamma_0\Phi e^\alpha{}_\beta\zeta^\beta
\\
\delta\hat{\cal{C}}^\alpha&=&\frac{8a_0\beta_0}{c_0}{\cal{B}}^{\alpha\beta}\zeta_\beta
+\frac{b_0\ti\delta_0}{c_0}\Pi^{\alpha\beta}\zeta_\beta
+\rho_0\Phi\zeta^\alpha\\
\delta\hat{\cal{C}}^{\alpha(2k+1)}&=&\tilde\beta_k{\cal{B}}^{\alpha(2k+1)\beta}\zeta_\beta
+\tilde\alpha_k
{\cal{B}}^{\alpha(2k)}\zeta^\alpha+\tilde\delta_k\Pi^{\alpha(2k+1)\beta}\zeta_\beta
+\tilde\gamma_k\Pi^{\alpha(2k)}\zeta^\alpha\\
\end{eqnarray*}
one gets
\begin{eqnarray*}
\gamma_{k}{}^2&=& \frac{(s-k-1)}{k(k+1)^2(2k+1)^2}[ M-(k+1) \lambda
]\hat\gamma^2
\\
\delta_{k}{}^2&=&\frac{(s+k+1)}{(k+1)(k+2)(2k+3)}[
M+(k+1)\lambda]\hat\delta^2
\end{eqnarray*}
$$
\tilde\gamma_k=\gamma_k,\quad\tilde\delta_k=\delta_k,\quad
\tilde\alpha_k=\frac{\alpha_k}{(2k+1)},\quad\tilde\beta_k=2(k+1)\beta_{k+1}
$$
$$
\gamma_0=-2\ti\gamma_0=\frac{c_0{}^2}{2a_0}\beta_1,\quad
\rho_0=-\frac{c_0{}^2}{4sMa_0}\beta_1
$$
where
$$
\hat\gamma=\frac{sM}{2}\hat\alpha,\qquad\hat\delta=\frac{sM}{\sqrt2}\hat\alpha
$$
Supersymmetric Lagrangian have form (\ref{LagSC_s}) and it is
invariant under the supertransformations up to equations of motion for
auxiliary fields $B^{\alpha(2)},\pi^{\alpha(2)}$ (\ref{Eq_low}).

\section{Realization of $AdS_3$ (super)algebra}
In this section we analyze the commutators of (super)transformations
and show how the (super)algebra is realized in our construction. All
considerations are valid both for $(s,s+1/2)$ supermultiplets and
for $(s,s-1/2)$ one.

\subsection{ $AdS$-transformations}

Before we turn to supersymmetric theory let us discuss the
conventional $AdS_3$ algebra. In the frame formalism, $AdS$ space is
described by background Lorentz connection fields
$\omega^{\alpha(2)}$ and background frame field $e^{\alpha(2)}$.
First of them enters implicitly through the covariant derivative
$D$, second one enters explicitly. Let $\eta^{\alpha(2)}$ and
$\xi^{\alpha(2)}$ be the parameters of Lorentz transformations and
pseudo-translation respectively. The theory of massive spin-$s$
field has the following laws under these transformations
\begin{eqnarray}\label{Lorentz_b1}
\delta_\eta\Omega^{\alpha(2k)}=\eta^\alpha{}_\beta\Omega^{\alpha(2k-1)\beta}\qquad
\delta_\eta f^{\alpha(2k)}=\eta^\alpha{}_\beta f^{\alpha(2k-1)\beta}
\end{eqnarray}
\begin{eqnarray}\label{Translaton_b1}
\delta_\xi\Omega^{\alpha(2k)} &=& \frac{(k+2)a_k}{k} \xi_{\beta(2)}
\Omega^{\alpha(2k)\beta(2)} + \frac{a_{k-1}}{k(2k-1)}
\xi^{\alpha(2)} \Omega^{\alpha(2k-2)}
 + \frac{b_k}{k} \xi^\alpha{}_\beta f^{\alpha(2k-1)\beta} \nonumber \\
\delta_\xi f^{\alpha(2k)} &=& \xi^\alpha{}_\beta
\Omega^{\alpha(2k-1)\beta} + a_k \xi_{\beta(2)}
f^{\alpha(2k)\beta(2)} + \frac{(k+1)a_{k-1}}{k(k-1)(2k-1)}
\xi^{\alpha(2)} f^{\alpha(2k-2)}
\end{eqnarray}
here $a_k$ and $b_k$ are defined by (\ref{boson_data}). For massive
spin-$(s\pm1/2)$ field the transformation laws look like
\begin{eqnarray}\label{ADStr_f}
\delta_\eta\Phi^{\alpha(2k+1)} &=&
\eta^\alpha{}_\beta\Phi^{\alpha(2k)\beta}
\nonumber \\
\delta_\xi\Phi^{\alpha(2k+1)} &=& \frac{d_k}{(2k+1)}
\xi^\alpha{}_\beta \Phi^{\alpha(2k)\beta}  + \frac{c_k}{k(2k+1)}
\xi^{\alpha(2)} \Phi^{\alpha(2k-1)}\nonumber\\
&&
 + c_{k+1} \xi_{\beta(2)} \Phi^{\alpha(2k+1)\beta(2)}
\end{eqnarray}
here $c_k$ and $d_k$ are defined by (\ref{fermion_data}) for
spin-$(s+1/2)$ and (\ref{InD}) for spin-$(s-1/2)$. To consider a
structure of the $AdS_3$ algebra $Sp(2) \otimes Sp(2)$ only in left
sector we introduce the new variables for bosonic fields
\begin{eqnarray}\label{NV}
\hat\Omega^{\alpha(2k)}=\Omega^{\alpha(2k)}+\frac{sM}{2k(k+1)}f^{\alpha(2k)},\qquad
\hat
f^{\alpha(2k)}=\Omega^{\alpha(2k)}-\frac{sM}{2k(k+1)}f^{\alpha(2k)}
\end{eqnarray}
so that the variables $\hat\Omega^{\alpha(2k)}$ correspond to left
sector. In terms of this variables the transformations
(\ref{Lorentz_b1}), (\ref{Translaton_b1}) have form
\begin{eqnarray}\label{ADStr_b}
\delta_\eta\hat\Omega^{\alpha(2k)}&=&\eta^\alpha{}_\beta\hat\Omega^{\alpha(2k-1)\beta}
\nonumber\\
\delta_\xi\hat\Omega^{\alpha(2k)} &=& \frac{(k+2)a_k}{k}
\xi_{\beta(2)} \hat\Omega^{\alpha(2k)\beta(2)} +
\frac{a_{k-1}}{k(2k-1)} \xi^{\alpha(2)}
\hat\Omega^{\alpha(2k-2)}\nonumber\\
&&
 + \frac{sM}{2k(k+1)} \xi^\alpha{}_\beta
\hat\Omega^{\alpha(2k-1)\beta}
\end{eqnarray}
Now let us consider the commutators of these transformations. The
direct calculations lead to the following results
\begin{eqnarray*}
{[\delta_{\eta_1},\delta_{\eta_2}]}\hat\Omega^{\alpha(2k)}&=&(\eta_2{}^\alpha{}_\beta\eta_1{}^\beta{}_\gamma-
\eta_1{}^\alpha{}_\beta\eta_2{}^\beta{}_\gamma)\hat\Omega^{\alpha(2k-1)\gamma},
\\
{[\delta_{\xi_1},\delta_{\xi_2}]}\hat\Omega^{\alpha(2k)}&=&
\frac{\lambda^2}{4} (\xi_2{}^\alpha{}_\gamma\xi_1{}^\gamma{}_\beta-
\xi_1{}^\alpha{}_\gamma\xi_2{}^\gamma{}_\beta)
\hat\Omega^{\alpha(2k-1)\beta},
\\
{[\delta_\eta,\delta_\xi]}\hat\Omega^{\alpha(2k)}&=&2\frac{(k+2)a_k}{k}
\xi_{\beta(2)}\eta^\beta{}_\gamma\hat\Omega^{\alpha(2k)\beta\gamma}
 + \frac{a_{k-1}}{k(2k-1)}\xi^{\alpha}{}_{\gamma}\eta^\alpha{}^\gamma
\hat\Omega^{\alpha(2k-2)}\\
&&  + \frac{sM}{2k(k+1)} (\xi^\alpha{}_\beta
\eta^\beta{}_\gamma-\eta^\alpha{}_\gamma\xi^\gamma{}_\beta)\hat\Omega^{\alpha(2k-1)}
\end{eqnarray*}
Comparison with (\ref{ADStr_b}) shows that we have the $AdS$-algebra
$$
[M_{\alpha(2)},M_{\beta(2)}]\sim\varepsilon_{\alpha\beta}M_{\alpha\beta}
,\qquad
[P_{\alpha(2)},P_{\beta(2)}]\sim\lambda^2\varepsilon_{\alpha\beta}M_{\alpha\beta}
$$
$$
[M_{\alpha(2)},P_{\beta(2)}]\sim\varepsilon_{\alpha\beta}P_{\alpha\beta}
$$
An analogous results have place for the commutators in the fermionic
sector.

\subsection{ Supertransformations}

Let us consider the supersymmetric theory.  The supertransformations
for massive higher spin supermultiplets have form (\ref{Super_b}),
(\ref{Super_f})
\begin{eqnarray*}
\delta\Omega^{\alpha(2k)} &=&
\frac{isM}{2k(k+1)}\beta_k\Phi^{\alpha(2k-1)}\zeta^\alpha+\frac{isM}{2k(k+1)}\alpha_k\Phi^{\alpha(2k)\beta}\zeta_\beta
\\
\delta f^{\alpha(2k)} &=& i\beta_k\Phi^{\alpha(2k-1)}\zeta^\alpha
+i\alpha_k\Phi^{\alpha(2k)\beta}\zeta_\beta
\\
\delta\Phi^{\alpha(2k+1)} &=&
\frac{\alpha_k}{(2k+1)}\Omega^{\alpha(2k)}\zeta^\alpha+
2(k+1)\beta_{k+1}\Omega^{\alpha(2k+1)\beta}\zeta_\beta\\
&& +\frac{sM}{2k(k+1)(2k+1)}\alpha_kf^{\alpha(2k)}\zeta^\alpha+
\frac{sM}{(k+2)}\beta_{k+1}f^{\alpha(2k+1)\beta}\zeta_\beta
\end{eqnarray*}
where the parameters $\alpha_k$ and $\beta_k$ are defined by
(\ref{solution1}) for $(s,s+1/2)$ supermultiplets and
(\ref{SupMult2}) for $(s,s-1/2)$. In terms of new variables
(\ref{NV}) the supertransformations look like
\begin{eqnarray*}
\delta\hat\Omega^{\alpha(2k)} &=&
\frac{isM}{k(k+1)}\beta_k\Phi^{\alpha(2k-1)}\zeta^\alpha+\frac{isM}{k(k+1)}\alpha_k\Phi^{\alpha(2k)\beta}\zeta_\beta
\\
\delta \hat f^{\alpha(2k)} &=&0
\\
\delta\Phi^{\alpha(2k+1)} &=&
\frac{\alpha_k}{(2k+1)}\hat\Omega^{\alpha(2k)}\zeta^\alpha+
2(k+1)\beta_{k+1}\hat\Omega^{\alpha(2k+1)\beta}\zeta_\beta
\end{eqnarray*}
One can see that the $\hat f^{\alpha(2k)}$ fields are inert under
the  supertransformations. It just means that we have (1,0)
supersymmetry. Let us calculate the commutator of two
supertransformations. We obtain
\begin{eqnarray*}
[\delta_1,\delta_2]\hat\Omega^{\alpha(2k)}&=&isM\hat\alpha^2[
\frac{a_{k-1}}{k(2k-1)}\hat\Omega^{\alpha(2k-1)}\zeta_1{}^\alpha\zeta_2{}^\alpha+
\frac{2(k+2)a_k}{k}\hat\Omega^{\alpha(2k)\gamma\beta}\zeta_1{}_\beta\zeta_2{}_\gamma\\
&&
+\frac{sM}{k(k+1)}\hat\Omega^{\alpha(2k-1)\gamma}\zeta_1{}^\alpha\zeta_2{}_\gamma
+\lambda\hat\Omega^{\alpha(2k-1)\gamma}\zeta_1{}^\alpha\zeta_2{}_\gamma]
-(1\leftrightarrow2)
\end{eqnarray*}
\begin{eqnarray*}
[\delta_1,\delta_2]\hat\Phi^{\alpha(2k+1)} &=&
isM\hat\alpha^2[\frac{c_k}{k(2k+1)}\Phi^{\alpha(2k-1)}\zeta_1{}^\alpha\zeta_2{}^\alpha
+2c_{k+1}\Phi^{\alpha(2k+1)\gamma\beta}\zeta_1{}_\beta\zeta_2{}_\gamma\\
&&+
\frac{2d_k}{(2k+1)}\Phi^{\alpha(2k)\gamma}\zeta_1^\alpha\zeta_2{}_\gamma
+\lambda\Phi^{\alpha(2k)\gamma}\zeta_1^\alpha\zeta_2{}_\gamma ]
-(1\leftrightarrow2)
\end{eqnarray*}
Here we use the explicit expressions for $\alpha_k$ and $\beta_k$
and conditions
$$
\frac{2\beta_k{}^2}{(k+1)}+\frac{\alpha_k{}^2}{k(k+1)(2k+1)}=\hat\alpha^2[\frac{sM}{k(k+1)}+\lambda]
$$
$$
\frac{2\beta_{k+1}{}^2}{(k+2)}+\frac{\alpha_k{}^2}{k(k+1)(2k+1)}=
\hat\alpha^2[\frac{2d_k}{(2k+1)}+\lambda]
$$
$$
\alpha_{k-1}\beta_k=(k+1)\hat\alpha^2a_{k-1},\quad
\alpha_k\beta_{k+1}=(k+2)\hat\alpha^2a_k,\quad
\alpha_k\beta_k=(k+1)c_k\hat\alpha^2
$$
Comparing the commutators of supertransformations with
(\ref{ADStr_f}) we obtain the (1,0) $AdS_3$ superalgebra with the
commutation relation
\begin{eqnarray}\label{algebra}
\{Q_\alpha,Q_\beta\}\sim
P_{\alpha\beta}+\frac{\lambda}{2}M_{\alpha\beta}
\end{eqnarray}

As we see, the algebra of the supertransformations is closed. It is
worth emphasizing that we did not apply the equations of motion to
obtain the relation (\ref{algebra}) both in bosonic and in fermionic
sectors. This situation is analogous to one for massless higher-spin
fields in the three-dimensional frame-like formalism. Recall that in
the massive supermultiplets case the invariance of the Lagrangians is
achieved up to the terms proportional to the spin-1 and spin-0
auxiliary fields equations only. Note that in dimensions $d \ge 4$ one
would have to use equations for the higher spins auxiliary fields
as well (though in odd dimensions there exist examples of the theories
where Lagrangians are invariant without any use of e.o.m
\cite{Zan05}). The difference here comes from the well known fact that
all massless higher spin fields in three dimensions do not have any
local degrees of freedom.

\section{Summary}
Let us summarize the results. In this paper we have constructed the
Lagrangian formulation for massive higher spin supermultiplets in
the $AdS_3$ space in the case of minimal (1,0) supersymmetry. For
description of the massive higher spin fields we have adapted for
massive fields in three dimensions the frame-like gauge invariant
formalism and technique of gauge invariant curvatures. The
supersymmetrization is achieved by deformation of the curvatures by
background gravitino field and hence the supersymmetric Lagrangians
are formulated with help of background fields of three-dimensional
supergravity. In $AdS_3$ the space the minimal (1,0) supersymmetry
combines the massive fields in supermultiplets with one bosonic
degree of freedom and one fermionic one. As a result we have derived
the supersymmetric and gauge invariant Lagrangians for massive
higher spin $(s,s+1/2)$ and $(s,s-1/2)$ supermultiplets.

\section*{Acknowledgments}
I.L.B and T.V.S are grateful to the RFBR grant, project No.
15-02-03594-a for partial support. Their research was also supported
in parts by Russian Ministry of Education and Science, project No.
3.1386.2017.

\section{Appendix A. Massless bosonic fields}\label{massless_bosons}
In this
Appendix we consider the frame-like formulation for the massless
bosonic fields in three dimensional flat space. For every spin we
present field variables and write out the corresponding Lagrangian.
All massless fields with spin $s\geq1$ are gauge ones so that we also
present the gauge transformations for them.
\\
\\
{\bf Spin 0}
\\
It is described by physical 0-form $\varphi$ and auxiliary 0-form
$\pi^{\alpha(2)}$. Lagrangian looks like
\begin{eqnarray*}
{\cal L} &=& - E \pi_{\alpha\beta} \pi^{\alpha\beta} +
\pi_{\alpha\beta} E^{\alpha\beta} D \varphi
\end{eqnarray*}
\\
{\bf Spin 1}
\\
It is described by physical 1-form $A$ and auxiliary 0-form
$B^{\alpha(2)}$. Lagrangian looks like
\begin{eqnarray*}
{\cal L} &=& E B_{\alpha\beta} B^{\alpha\beta} - B_{\alpha\beta}
e^{\alpha\beta} D A
\end{eqnarray*}
It is invariant under gauge transformations
$$
\delta A = D \xi
$$
\\
{\bf Spin 2}
\\
It is described by physical 1-form $f^{\alpha(2)}$ and auxiliary
1-form $\Omega^{\alpha(2)}$. Lagrangian looks like
\begin{eqnarray*}
{\cal L} &=& \Omega_{\alpha\beta} e^\beta{}_\gamma
\Omega^{\alpha\gamma} + \Omega_{\alpha(2k)} D f^{\alpha(2k)}
\end{eqnarray*}
The gauge transformations have the form
$$
\delta \Omega^{\alpha(2)} = D \eta^{\alpha(2)},\qquad \delta
f^{\alpha(2)} = D \xi^{\alpha(2)} + e^\alpha{}_\gamma
\eta^{\alpha\gamma}
$$
\\
{\bf Spin $k$}
\\
It is described by physical 1-form $f^{\alpha(2k-2)}$ and auxiliary
1-form $\Omega^{\alpha(2k-2)}$. Lagrangian looks like
\begin{eqnarray*}
{\cal L} &=&  (-1)^{k+1} [ k \Omega_{\alpha(2k-1)\beta}
e^\beta{}_\gamma \Omega^{\alpha(2k-1)\gamma} + \Omega_{\alpha(2k)} D
f^{\alpha(2k)}
 ]
\end{eqnarray*}
The gauge transformations have the form
$$
\delta \Omega^{\alpha(2k)} = D \eta^{\alpha(2k)},\qquad \delta
f^{\alpha(2k)} = D \xi^{\alpha(2k)} + e^\alpha{}_\beta
\eta^{\alpha(2k-1)\beta}
$$

\section{Appendix B. Massless fermionic
fields}\label{massless_fermions} In
this Appendix we consider the frame-like formulation for the
massless fermionic fields in three dimensional flat space. For every
spin we present field variables and write out the corresponding
Lagrangian. All massless fields with spin $s\geq3/2$ are gauge ones so
that we also present gauge transformations for them.
\\
\\
{\bf Spin 1/2}
\\
It is described by master 0-form $\phi^\alpha$. Lagrangian looks
like
\begin{eqnarray*}
{\cal L} &=& \frac{1}{2} \phi_\alpha E^\alpha{}_\beta D \phi^\beta
\end{eqnarray*}
\\
{\bf Spin 3/2}
\\
It is described by physical 1-form $\Phi^\alpha$. Lagrangian and gauge
transformations have the form
\begin{eqnarray*}
{\cal L} &=& -\frac{i}{2} \Phi_{\alpha} D \Phi^{\alpha},\qquad
\delta \Phi^{\alpha} = D \xi^{\alpha}
\end{eqnarray*}
\\
{\bf Spin $k+1/2$}
\\
It is described by physical 1-form $\Phi^{\alpha(2k-1)}$. Lagrangian
and gauge transformations have the form
\begin{eqnarray*}
{\cal L} &=&  (-1)^{k+1}\frac{i}{2}  \Phi_{\alpha(2k+1)} D
\Phi^{\alpha(2k+1)},\qquad \delta \Phi^{\alpha(2k+1)} = D
\xi^{\alpha(2k+1)}
\end{eqnarray*}


\end{document}